\documentclass{aastex631}

\usepackage{xcolor}
\usepackage{amsmath}
\usepackage{ragged2e}
\usepackage{makecell}
\usepackage{float}  
\usepackage{subfigure}
\usepackage{hyperref}

\begin{document}

\title{Gamma rays as a signature of $r$-process producing supernovae: remnants and future Galactic explosions}

\correspondingauthor{Zhenghai Liu, Xilu Wang}
\email{zliu75@ncsu.edu, wangxl@ihep.ac.cn}

\author[0000-0002-8056-2526]{Zhenghai Liu}
\affiliation{Department of Physics, North Carolina State University, Raleigh, NC 27695, USA}

\author[0000-0002-9409-3468]{Evan Grohs}
\affiliation{Department of Physics, North Carolina State University, Raleigh, NC 27695, USA}

\author[0000-0003-0031-1397]{Kelsey A.\ Lund}
\affiliation{Department of Physics, University of California, Berkeley,  CA, USA}
\affiliation{Institute for Nuclear Theory, University of Washington, Seattle, WA, USA}

\author[0000-0001-6811-6657]{G. C. McLaughlin}
\affiliation{Department of Physics, North Carolina State University, Raleigh, NC 27695, USA}

\author[0000-0001-6653-7538]{M. Reichert}
\affiliation{Departament d’Astronomia i Astrofísica, Universitat de València, Edifici d’Investigació Jeroni Munyoz, C/Dr. Moliner, 50, E-46100 Burjassot (València), Spain}

\author[0000-0001-5107-8930]{Ian U.\ Roederer}
\affiliation{Department of Physics, North Carolina State University, Raleigh, NC 27695, USA}

\author[0000-0002-4729-8823]{Rebecca Surman}
\affiliation{Department of Physics and Astronomy, University of Notre Dame, Notre Dame, IN 46556, USA}

\author[0000-0002-5901-9879]{Xilu Wang}
\affiliation{State Key Laboratory of Particle Astrophysics, Institute of High Energy Physics, Chinese Academy of Sciences, Beijing, 100049, China}

\begin{abstract}

We consider the question of whether core-collapse supernovae (CCSNe) can produce rapid neutron capture process ($r$-process) elements and how future MeV gamma-ray observations could address this.  Rare types of CCSNe characterized by substantial magnetic fields and rotation, known as magnetorotational supernovae (MR-SNe), are theoretically predicted to produce these elements, although direct observational evidence is lacking.  We suggest that this critical question be addressed through the study of some of the eleven CCSN remnants located within 10 kpc, as well as through the detection of gamma-ray emission from a future Galactic supernova. We use a two-dimensional MR-SN model to estimate the expected gamma flux stemming from nuclear decays in the range of a few tens of keV to a few MeV.  Our results indicate that an observation of $^{126} {\rm Sn}$ ($^{126}$Sb) in a remnant stands out as a signature of an $r$-process-producing supernova.  Since the neutron-rich conditions that lead to the production of the $r$-process could also enhance the production of $^{60}{\rm Fe}$, the detection of substantial $^{60}{\rm Fe}$ ($^{60}{\rm Co}$)  would be indicative of favorable conditions for the $r$-process.  In the case of a future supernova explosion, when the evolution of the spectrum is studied over ten days to a few years, a rich picture emerges.  At various epochs, second peak $r$-process isotopes such as $^{125}{\rm Sn}$, $^{131}{\rm I}$, $^{131}{\rm Te}$, $^{132}{\rm I}$ and $^{140}{\rm La}$ produce gamma-ray signals that emerge above the background from explosive burning products and electron–positron annihilation.
The weak $r$-process isotopes $^{95}{\rm Nb}$, $^{103}{\rm Ru}$, $^{106}{\rm Rh}$ also have periods of prominence.  While MR-SNe are predicted to have a relatively small main $r$-process contribution, third peak isotopes like $^{194}{\rm Ir}$ could still be above next generation MeV gamma instrument sensitivities.  

\end{abstract}

%
%

\section{Introduction}

Despite intense effort, a complete picture of the astrophysical locations of the rapid neutron capture process ($r$-process) of nucleosynthesis remains unsettled; for reviews, see \cite{RevModPhys.93.015002,2023A&ARv..31....1A}, and references within. 
Several different scenarios have been suggested, with compact object mergers being the most widely explored. Binary neutron star mergers have been extensively studied; for a review, see \citet{Radice:2020ddv}. Black hole–neutron star mergers have also been the focus of various efforts (e.g., \citealt{Surman:2008qf,2015PhRvD..91l4021F,2017MNRAS.464.3907R,Curtis:2022hjy,2023ApJ...943L..12K}).  Besides mergers, the explosions of massive stars are also possible scenarios. While theoretical models of neutrino/advection driven core collapse supernovae (CCSNe) currently do not produce $r$-process elements, CCSNe that explode via other mechanisms may do.  
For example, collapsars are a scenario where the collapse of a rotating massive star leaves a black hole-accretion disk remnant system \citep{Popham:1998ab,MacFadyen:1998vz};  accretion disk winds could provide the conditions to form $r$-process elements \citep{Surman:2005kf,Siegel:2018zxq,Miller2020_collapsar, Agarwal2025_supercollapsar}. Another possible outcome is a magnetorotational supernova (MR-SN), where a strong magnetic field assists in the formation of a jet along the rotational axis that ejects neutron rich material \citep{2015ApJ...810..109N, 2017ApJ...836L..21N, 2018ApJ...864..171M,2021MNRAS.501.5733R, Obergaulinger2023, 2023MNRAS.518.1557R}. Additional proposed $r$-process sites include primordial black hole-neutron star destruction \citep{Fuller:2017uyd}, neutrino-driven processes in the helium shell of CCSNe \citep{Banerjee:2011zm}, and magnetar giant flares \citep{Patel:2025frn}.  

In addition to theoretical predictions, observations are crucial to our understanding of the Galactic inventory of heavy elements. The strongest observational evidence for the in situ synthesis of $r$-process elements comes from the observation of kilonovae, which are the electromagnetic counterparts to binary neutron star mergers. In kilonovae, optical/infrared emission appears to be redder and decays more slowly when lanthanides are present in the ejecta.  Since lanthanides are produced via $r$-process nucleosynthesis, this indicates at least a partial $r$-process \citep{2013ApJ...774L..23B,2017LRR....20....3M,2017Natur.551...80K,2021ApJ...918...44B}. In one binary neutron star merger,  both the gravitational wave signal, GW170817, and the electromagnetic counterpart were detected \citep{2017ApJ...848L..12A,2017PhRvL.119p1101A,2017ApJ...848L..21A,2017ApJ...848L..17C,2017ApJ...851L..21V} and a line from the weak $r$-process element Sr was observed \citep{Watson:2019xjv}.    
An alternative to in situ observations is the detection of heavy-element signatures in metal-poor stars, e.g. \cite{2008ApJ...675..723R, 2012ApJ...750...76R}; also see \cite{2018ARNPS..68..237F} for a review. 
Observations of metal-poor stars in dwarf galaxies suggest that early $r$-process formation comes from a relatively rare event \citep{Ji:2015wzg, 2016AJ....151...82R}, which could be, for example, binary neutron star mergers, black hole-neutron star mergers, collapsars or MR-SNe.  Data from metal poor stars has also been used to suggest that rare types of CCSNe may be an important contributor to Galactic $r$-process, perhaps even moreso than compact object merger scenarios \citep{2019A&A...631A.171S,2019ApJ...875..106C, 2020MNRAS.494.4867V, 2021Natur.595..223Y}. 
Additional information comes from the measurement of radioactive $r$-process nuclei on Earth, such as $^{244}$Pu, \citep{2021Sci...372..742W,2021APS..APRD09007W,2023ARNPS..73..365F,2023ApJ...948..113W,2025PSJ.....6...75B}  which provides a method to probe the time since production. 

A complementary observational approach is the detection of gamma rays produced by the beta decay of freshly synthesized radioactive isotopes in the days to years after an explosion. Gamma rays with energies $E_\gamma \lesssim$ a few MeV carry isotope-specific signatures and offer a powerful probe of nucleosynthesis processes and yields.  Up to half of the radioactive energy of a kilonova produced on timescales of hours to days postexplosion is released in gamma-rays. This gamma-ray emission is potentially detectable if the event occurs in the Galaxy \citep{Hotokezaka2016}. Specific isotopes produced in kilonovae could be identified with future detectors \citep{2019ApJ...872...19L,2020ApJ...889..168K,2021ApJ...919...59C} with key signatures including gamma rays from fissioning nuclei which dominate above 3.5 MeV \citep{2020ApJ...903L...3W},
and the 2.6 MeV line of $^{208}$Tl  \citep{Vassh:2023fnb} which would indicate that the $r$-process proceeded past the third peak.  Alternative $r$-process sites, such as magnetar giant flares, could also produce MeV gamma emission \citep{2025ApJ...984L..29P} detectable by future detectors.

Due to the uncertainty about the primary origin of the Galactic $r$-process elements, direct observational evidence that points to the existence of an $r$-process from supernovae would be welcome. 
Three decades ago, at a time when it was thought that a robust $r$-process came from neutrino-driven supernovae \citep{1992ApJ...399..656M,1994ApJ...433..229W,1994A&A...286..857T}, \citet{Qian:1998cz} highlighted several longer-lived $r$-process isotopes as promising candidates for gamma-ray detection from future Galactic CCSNe, although they considered this detection to be challenging without a highly sensitive gamma detector. We revisit this scenario in the current context, where collapsars and MR-SNe are considered as potential $r$-process sources, and instruments have greatly improved sensitivities. We use the specific example of an MR-SN as a proxy for $r$-process-producing supernovae. Although MR-SNe may produce a less robust $r$-process compared to neutron star mergers \citep{2021MNRAS.501.5733R}, their high-energy ejecta synthesize a range of radioactive nuclei up to and beyond the second peak. There are two detection avenues to consider: the prompt signal, which lasts from days to a few years, and the signal from supernova remnants (SNRs).

Supernova remnants offer unique laboratories for gamma-ray observations of radioactive isotopes produced during stellar explosions due to their extended ages.  They are typically hundreds to thousands of years old, and if the synthesis of the $r$-process occurred in the supernova, then long-lived isotopes which are unstable to beta decay should be present in the remnant and emit both x-rays \citep{2014MNRAS.438.3243R} and gamma rays \citep{Qian:1998cz}. Furthermore, many Galactic remnants are located at relatively close distances, typically within 10 kpc, which enhances the prospects for detecting radioactive ejecta. As we shall discuss, in MR-SNe the predicted yields of some long-lived $r$-process isotopes, such as $^{126}{\rm Sb}$, can be similar to those of the commonly studied $^{44}{\rm Ti}$, although smaller than the yield of $^{56}{\rm Co}$. In addition, the predicted yields of $^{60}{\rm Fe}$ can be {\it larger} than those of $^{44}{\rm Ti}$. 
As  $^{44}{\rm Ti}$ has already been detected in remnants a number of times, for example in Cassiopeia A by INTEGRAL/SPI 
\citep{2015A&A...579A.124S,2020A&A...638A..83W}, NuSTAR \citep{2014Grefenstette,2017ApJ...834...19G}, and INTEGRAL IBIS/ISGRI \citep{2006Renaud},  
and $^{60}{\rm Fe}$  has been detected as diffuse Galactic emission \citep{2007A&A...469.1005W}, it is timely to consider the possibility of discovering neutron-rich isotopes from SNRs.

The feasibility of detecting neutron-rich isotopes from supernova remnants or future Galactic supernovae depends on the sensitivity of gamma-ray instruments. The Spectrometer on INTEGRAL (SPI), launched by the European Space Agency in 2002 \citep{2003A&A...411L..91R}, had been the primary observatory for MeV-range gamma rays until it ceased data collection on February 28, 2025. For future MeV detectors, the Compton Spectrometer and Imager (COSI) is scheduled to launch and begin observations in 2027 \citep{Tomsick:2019wvo,Tomsick:2023aue}, and promises improved sensitivity relative to SPI.  
Furthermore, several next-generation gamma-ray detectors have also been proposed in recent years, including observatory-scale missions such as AMEGO\citep{McEnery2019All}, eAstrogam \citep{2018JHEAp..19....1D}, and MeVGRO \citep{MEVGRO}, as well as small- to mid-scale missions such as GRAMS \citep{2020APh...114..107A}, GECCO \citep{2022JCAP...07..036O}, COCOA\citep{SOLETI2025103135}, GammaTPC \citep{2024HEAD...2150005S}, ComPair \citep{2024HEAD...2150004K}, MASS \citep{mass} and MeVCUBE \citep{2022JCAP...08..013L}, with substantial improvement in sensitivity compared to SPI in the MeV energy range.

In this article, we consider supernova gamma-ray emission stemming from beta minus and beta plus decay of radioactive isotopes. We consider an MR-SN model from \citet{2021MNRAS.501.5733R},
identify characteristic signatures both for a hypothetical future MR-SN event and for existing SNRs, and compare with instrument sensitivities. 
In Section \ref{sec:method}, we briefly describe our methods and the underlying physical models. In Section \ref{sec:remnant}, we present results for SNRs. In Section \ref{sec:earlystage}, we consider a future Galactic MR-SN event and discuss the time evolution of the gamma-ray spectra.

%
%
%

\section{Method}
\label{sec:method}
\subsection{Element synthesis from the MR-SN model}
\label{sec:gammas}

\begin{figure}[htbp]
    \centering
     \centering
    \includegraphics[width=0.8\linewidth]{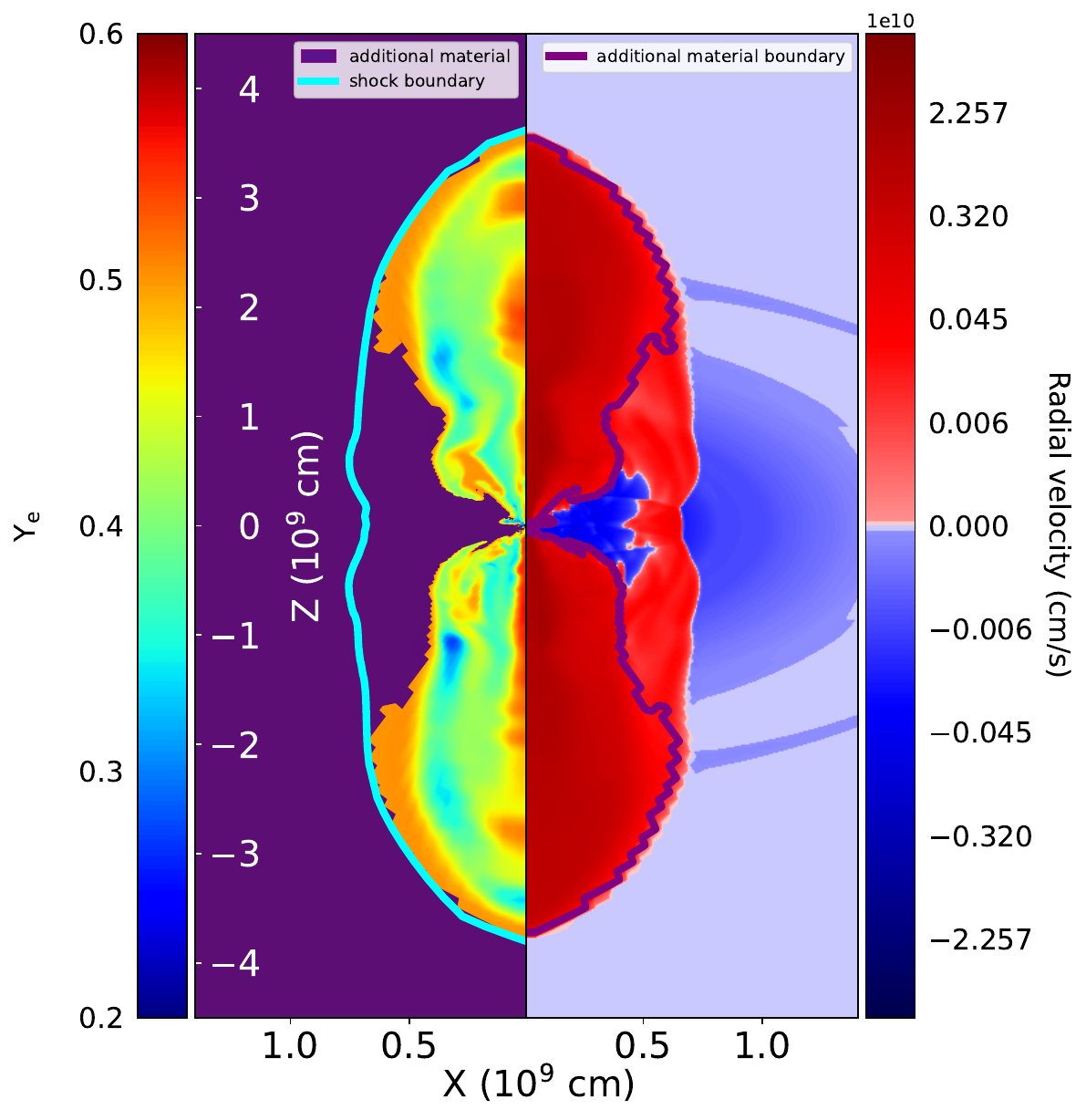}
    \caption{ Snapshot of the MR-SN model 35OC-Rs from \citet{2021MNRAS.501.5733R} at the end of the simulation time (1.306 s). \textbf{Left Panel}:  Tracers are drawn from the jet material in the multi-colored region;  the colormap corresponds to the electron fraction (${\rm Y}_{\rm e}$).  
    This material was unbound at the end of the simulation time; however, since the simulation ends before the explosion is complete, additional material (purple region, refer to the text for composition) is expected to be ejected by the end of the explosion.  The light blue line represents the outline of the outgoing shock. \textbf{Right Panel}: Red color indicates material with positive (outgoing) radial velocity while blue corresponds to negative (inward-going) radial velocity. The purple line denotes the boundary of the material in the jet at 1.3s and everything else (labeled as additional material).}
    \label{fig:snapshot}
\end{figure}

By using the initial conditions and 
thermodynamic properties of supernova ejecta
to perform nucleosynthesis calculations with a reaction network code, we can determine the time evolution of the ejecta composition and identify potential isotopic gamma-ray signatures. In this study, we utilize tracers from the 35OC-RS model, as described by \cite{2017MNRAS.469L..43O} and \cite{2021MNRAS.501.5733R}, which is based on a two-dimensional simulation of an MR-SN that incorporates neutrino transport within the M1 framework. 
This model produces fairly neutron-rich ejecta compared to other models in the same study, but has a neutron-richness in line with MR-SN models from other studies \citep{2015ApJ...810..109N,2017ApJ...836L..21N}. Due to its computational cost, the dynamical simulation ends  at 1.3~s. At that time, sufficient energy has been imparted to some of the material in the jet so that it has already become unbound. This area is illustrated by the multicolored region in the left panel of Fig. \ref{fig:snapshot}.  We anticipate that if the simulation were to run longer, additional material would be ejected by the jet, as it is behind the shock boundary and has insufficient angular momentum to halt its fall onto the protoneutron star \citep{2023MNRAS.518.1557R}.  This region is shown as the small blue region near the equator in the right panel of Fig. \ref{fig:snapshot}.  In all other regions, we assume shock-heated ejecta. 

%
%

The jet material
which has positive energy and outward radial velocity at the end of the simulation
is colored by the electron fraction (Y$_e$) in the left panel of Fig. \ref{fig:snapshot}.   
The total mass in this region is approximately 
0.4 M$_{\odot}$. Evolution trajectories for this region are calculated using Lagrangian tracer particles that are set up at the start of the simulation \citep{2021MNRAS.501.5733R}. 
To model their expansion, we evolve the nucleosynthesis according to the thermodynamic conditions in the tracer up to 1.3 s and then
extrapolate 
using an adiabatic expansion ($\rho \propto t^{-3}$). 
As all of these trajectories begin at sufficiently high temperatures that the composition can be represented by nuclear statistical equilibrium, the initial conditions for each trajectory are set by the SFHo equation of state \citep{Steiner_2013sfho}.

As mentioned earlier, we anticipate that the blue heart-shaped ``blob" of infalling material near the center in the right panel of Fig.~\ref{fig:snapshot} will accrete onto the proto-neutron star before being ejected in the jet. This material totals approximately 0.3 M$_{\odot}$. As indicated by \cite{2021MNRAS.501.5733R}, this infalling material is unlikely to be neutron-rich enough to contribute significantly to the main $r$-process.  Therefore, we assume its thermodynamic and hydrodynamic evolution mimics that of the jet ejecta, except for the neutron-rich main $r$-process producing material. 

We now turn to element synthesis in the shock-ejected outflow, which is very similar to the explosive burning that dominates the element synthesis in more common convective/neutrino-driven supernovae. 
We first consider the regions that have negative energy and inward radial velocity at the end of the simulation, shown as the outer blue region in the right panel of Fig. \ref{fig:snapshot}. This material totals approximately 20.5 M$_\odot$. We take the initial composition to be that of the progenitor star, following the approach of \cite{2000ApJ...528..368H}, which assumes minimal displacement from the pre-explosion state before the arrival of the shock. We use the methods described in \cite{2023MNRAS.518.1557R} and \cite{2002A&A...386..711N} to model the passage of the shock.  The shock initiates nuclear burning leading to nucleosynthesis outcomes typical of convective/neutrino-driven CCSNe. For example, the inner regions are the primary contributors to Fe-peak nucleosynthesis in the explosion, producing isotopes such as $^{56}$Ni and $^{56}$Co.  Decays from these isotopes dominate the early gamma-ray spectrum.  While we do not know for sure if the entire star would be ejected during the explosion, for the purposes of this study, we assume $100\%$ ejection, treating the resulting beta plus gamma spectrum as an upper limit. In the right panel of Fig. \ref{fig:snapshot} we see some material that has a positive radial velocity, shown in red outside the boundary that separates material already in the jet at 1.3s from everything else 
(labeled as additional material in the figure). Despite already having an outward velocity, this material, which contains a mass of roughly 1.5 M$_\odot$, has undergone minimal burning at the end of the simulation and is treated similarly to the other shock-ejected outflow. 

\begin{figure}[htbp]
    \centering
    \parbox{0.49\textwidth}{
        \centering
        \includegraphics[width=\linewidth]{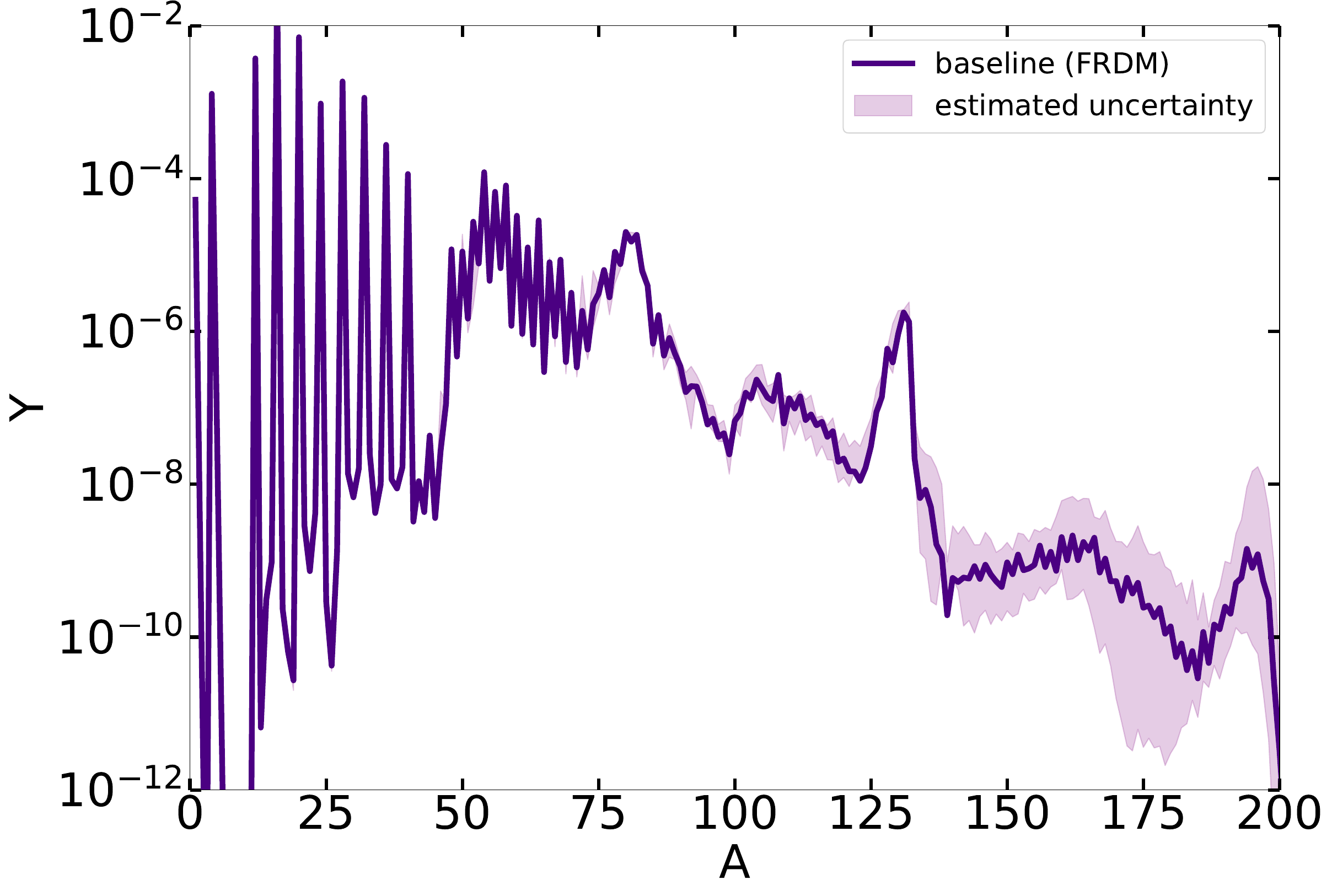}
    }
    \hfill
    \parbox{0.49\textwidth}{
        \centering
        \includegraphics[width=\linewidth]{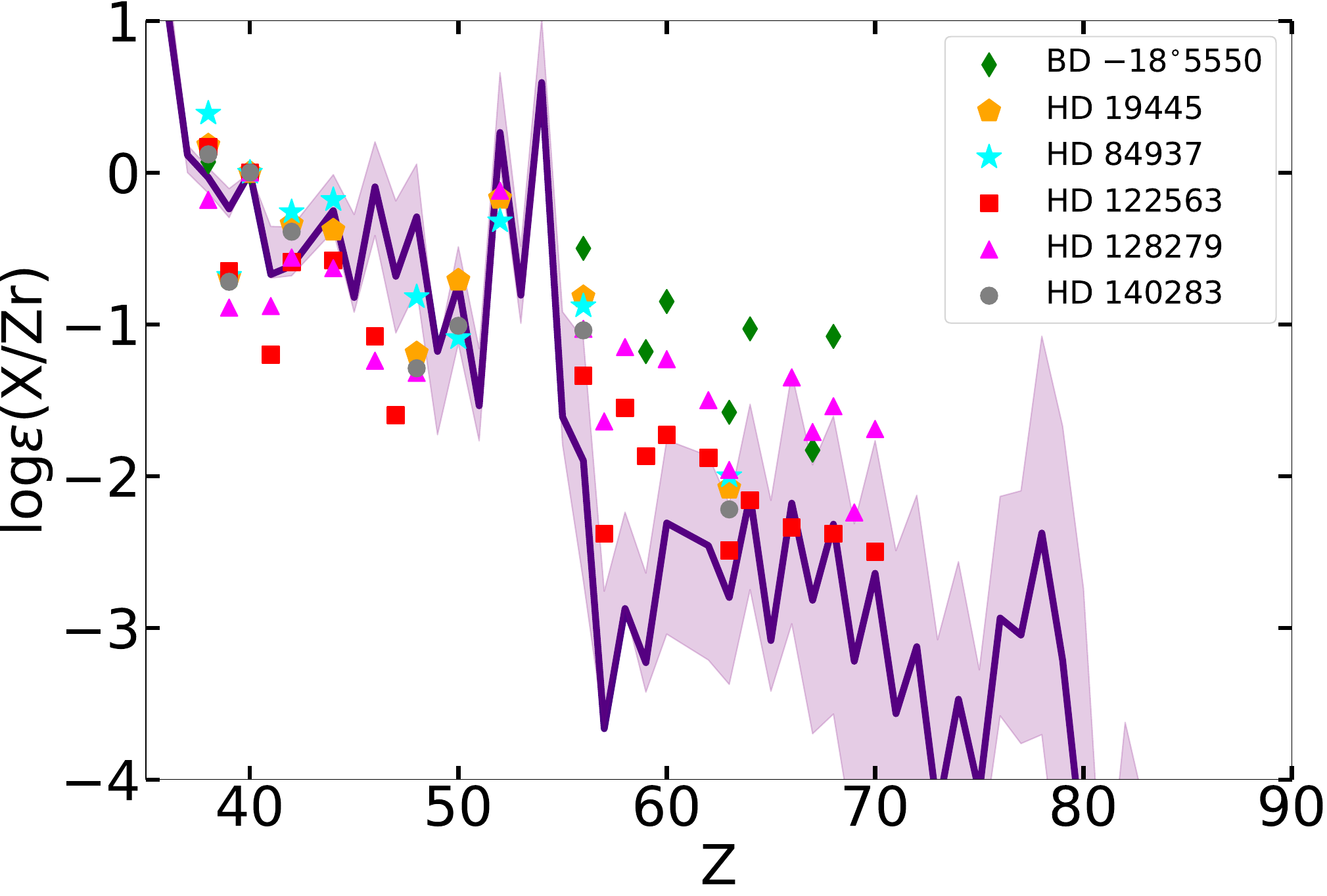}
    }
    \caption{Final abundances for ejecta from the model shown in Fig. \ref{fig:snapshot} plotted against mass number (left panel) and element number (right panel).
    The shaded region indicates the nuclear uncertainties introduced by different nuclear mass models and $\beta$ decay rates in the $r$-process region. 
    The abundances in the right panel are normalized to Zr ($Z$ = 40). Several observed stellar abundance patterns are shown for comparison: \mbox{HD~122563} \citep{2006honda}, \mbox{BD~$-$18$^{\circ}$5550} \citep{2004Cay} \citep{2007A&A...476..935F}, \mbox{HD~84937}, \mbox{HD~19445}, \mbox{HD~128279}, and \mbox{HD~140283} \citep{2022Ian}.}
    \label{fig:abundance}
\end{figure}

We utilize the PRISM nuclear reaction network from \cite{Sprouse+2021} to determine the nuclear abundances and reaction rates as a function of time for our model. 
Figure \ref{fig:abundance}  presents the final abundance distributions at 1 Gyr post-explosion. 
We pause briefly to consider uncertainties in our predictions of heavy nuclei introduced by unknown nuclear properties of neutron-rich nuclei heavier than iron. 
The shaded bands represent uncertainties introduced by variations in mass models FRDM \citep{FRDM}, HFB27 \citep{HFB27}, UNEDF1 \citep{unedf1,2015PhRvC..91b4310M}, DZ33 \citep{1995PhRvC..52...23D}, propagated to neutron capture rates and beta-decay $Q$ values as in \cite{Mumpower+2018}, and Reaclib+Nubase \citep{2010reaclib}. For this study, we used beta decay rates described in \cite{2016MKT} and \cite{2019ADNDT.125....1M} where experimental values \citep{Kondev_2021} are not available.   For the remainder of the study, we adopt the FRDM masses and M\"{o}ller beta decay rates as shown as the dark purple line in Fig. \ref{fig:abundance} as the representative case. 

In Fig. \ref{fig:abundance}, panel (a) shows the overall mass-number abundance,
while panel (b) compares the predicted elemental abundances with observational data from metal-poor stars, \mbox{HD~122563} \citep{2006honda}, \mbox{BD~$-$18$^{\circ}$5550} \citep{2007A&A...476..935F}, \mbox{HD~84937}, \mbox{HD~19445}, \mbox{HD~128279}, and \mbox{HD~140283} \citep{2022Ian}, normalized to zirconium (Zr, $Z$ = 40). The predicted abundance pattern agrees well with observational data in the weak $r$-process region ($Z$ $<$ 56), where the uncertainties from the nuclear inputs we consider remain relatively low. Two distinct peaks at tellurium (Te, $Z$ = 52) and xenon (Xe, $Z$ = 54) are seen in panel (b), consistent with the strong second peak seen in panel (a). The local Te maxima also appear in stellar data and are within one order of magnitude of our MR-SN model predictions, when normalized to the Zr abundance. Tellurium is the decay product of radioactive antimony (Sb, $Z$ = 51), which will be discussed in more detail later in the paper. 

In the main $r$-process region, predicted abundances are generally lower than those observed in metal-poor stars, except \mbox{HD~122563}.  This result is sensitive to the amount of very low ${\rm Y}_{\rm e}$ material and, therefore, also to our assumption that no additional $r$-process material past the second peak will be produced in addition to what is already in the jet at 1.3 s.  Future dynamical simulations that run for a longer time will be necessary to better understand if there is indeed a discrepancy with simulations in the main $r$-process region.

%
%
%

%
%

\subsection{Emitted gamma-ray spectra}

The emitted gamma-ray spectra can be calculated at each time
using the results from the PRISM reaction network, together with nuclear data on gamma decay intensities for each isotope. For the latter, we use the ENDF/B-VIII.0 database \citep{2018NDS...148....1B} and the ENSDF database (available at http://www.nndc.bnl.gov/ensarchivals/), both accessed in 2024. We compute the gamma spectrum for each trajectory; i.e., either tracer particles or components of the shock ejected outflow, 
as in \cite{2020ApJ...889..168K} and \cite{2020ApJ...903L...3W}, by using  
\begin{equation}
    \frac{dN^{prompt}_j}{dE_{rest}dt} (E_{rest}) = \left(\sum_{i}\text{flow}_{ij}\cdot P_{i}(E_{rest}) \right)
    \frac{N_{A} M_{j}}{dE_{rest}}
    \label{equation:dNdtdAdE}
\end{equation}
and then summing the gamma spectra over all trajectories.
In this equation, $N^{prompt}_j$ represents the number of gammas produced by nuclear beta decay in a given trajectory $j$, while 
\(\text{flow}_{ij} = \lambda_{i}Y_{ij}\) is the number of beta decays per second for isotope $i$,  $\lambda_{i}$ is the beta decay rate and $Y_{ij}$ is the abundance of isotope \( i \) in trajectory $j$ at the time when the flow is evaluated. For each beta decay, the gamma decay intensity for each energy channel is given by \( P_{i}(E_{rest}) \). Note that \( P_{i}(E_{rest}) \) is an intensity for gammas from a particular nuclear decay and is distinct from 
the line intensity discussed in the next section. Additional parameters include Avogadro’s number, $N_{A}$, the trajectory mass, $M_{j}$, and the energy of the gamma in the rest frame of the emmitting isotope, $E_{rest}$. 
In our calculations, we use an energy bin width, $dE$, of 0.01 MeV. Equation \ref{equation:dNdtdAdE} can be used for both beta minus and beta plus decays.
Although most nuclei we study have experimentally measured beta decay and gamma energy data, in the few exceptions that do not,
we use the theoretical continuum component available in the ENDF/B-VIII.0 database for $P_i(E_{rest})$. We neglect the possibility of anything other than decays from the ground state, leaving such an investigation to future work. 

%
%

\subsection{Annihilation}
\label{sec:annihilation}

Another important source of gamma-ray emission is 
the gamma radiation produced by positron--electron annihilation. In CCSNe, positrons are primarily generated through beta-plus decays of several nuclei, including \({}^{56}\)Ni, \(t_{1/2} \approx 6\) days, \({}^{56}\)Co, \(t_{1/2} \approx 77\) days, and \({}^{44}\)Sc, \(t_{1/2} \approx 4\) hours; the latter is the electron capture decay product of \({}^{44}\)Ti, \(t_{1/2} \approx 59.1\) years. In laboratory conditions, these decays are observed to produce 511\,keV photons with varying intensities: \({}^{56}\)Ni (\(P_{511} = 0.0013\%\)), \({}^{56}\)Co (\(P_{511} = 38\%\)), and \({}^{44}\)Sc (\(P_{511} = 188\%\)) \footnote{National Nuclear Data Center, information extracted from the NuDat database, https://www.nndc.bnl.gov/nudat/}. These nuclei either decay by beta plus decay or by electron capture, producing 511 keV intensities which are less than 200\%.

 Interstellar 511\,keV emission has been both predicted and observed \citep{2020NewAR..9001548C,McEnery2019All,2012A&A...543A...3M}, for a review, see \cite{Prantzos2010}. Similar annihilation radiation can also arise directly from the prompt emission of supernovae and young supernova remnants, though to date no 511 keV emission has been observed from a point source. In contrast to laboratory settings some positrons may escape the ejecta without annihilating. Therefore, it is important to estimate the in situ annihilation fraction to assess the contribution of the annihilation gammas to the observed gamma-ray spectrum.  Such radiation would be interesting to detect in its own right, although it may also act as a background that interferes with the detection of gammas that originate from nuclear decays in the same energy range. 

The transport of positrons through the ejecta is a complex process involving many interactions and depends sensitively on local density, composition, and ionization state as discussed in \citet{Prantzos2010}. Adopting the approximation of a uniformly mixed medium of constant electron number density $n_e$, we use the Bethe-Bloch equation to determine $l(n_e)$, the distance that a positron can travel before losing its kinetic energy.  If the positron is still within the ejecta after traveling a distance $l(n_e)$, we assume that it will annihilate in flight or form positronium (which subsequently decays).  If it is not within the ejecta after traveling a distance $l(n_e)$, we assume it will escape. This leads to an estimate of
its annihilation/decay probability $P(l(n_e))$,  
\begin{align}
  P_{\rm ann}(\ell) = 1 - \frac{\ell}{16R_{\text{max}}}\left[12-\left(\frac{\ell}{R_{\text{max}}}\right)^2\right],  \qquad
  0<\ell<2R_\text{max}
  \label{eq:annprob2}
\end{align}
 where $R_{\text{max}}$ is the radius of spherical ejecta at a given time; see Appendix \ref{app:annihilation} for details. For $l(n_e) > 2 R_{\text{max}}$, we assume that all positrons escape the ejecta.  
With the use of Eq. \ref{equation:dNdtdAdE} for beta plus decay flow, the electron number density in the medium,
and Eq. \ref{eq:annprob2}, we can estimate the rate of production of non-escaping positrons.  We make the approximation that all non-escaping positrons form positronium.  \textit{Para-positronium} (the singlet state) decays into two 511 keV photons, while \textit{ortho-positronium} (the triplet state) decays into three photons with a continuous energy distribution summing to 1022 keV.  The decay of positronium is via the singlet state 25\% of the time and via the triplet state 75\% of the time \citep{1951PhRv...82..455D, Prantzos2010, 2012A&A...543A...3M}.

\subsection{Absorption and Doppler broadening of the emitted gamma-ray photons}
\label{sec:doppler}

We calculate the final gamma-ray emission by propagating the emitted photons through the SN ejecta. To do so, we adopt an ejecta model, denoted as \lq uniform mix', that assumes homologous spherical expansion with a uniform density function and complete mixing of all ejecta components.  The expansion speed is taken to be $v (r) = v_{\text{ej}} r /R_{\text{max}} $ with the outmost layer speed being $v_{\text{ej}} = 0.015c$, and $R_{\rm max} (t)$ is the radius of the ejecta at time $t$.  
 At all times, rapid expansion of the ejecta induces Doppler shifts in the emitted gamma radiation, modifying the observed spectrum.  In addition, early-time gamma-ray spectra undergo significant reprocessing due to interactions such as Compton scattering, photoelectric absorption, pair production, etc.  To take account of these effects, we use a similar method outlined in \citet{2021APS..APRD09007W} and \citet{Vassh:2023fnb}. We briefly describe a second model with jet material distributed in an outer layer in Appendix \ref{app:radiationtransport}.  However, as it does not produce substantially different results than the \lq uniform mix' model, we do not present the results here. The details of these models are discussed in Appendix \ref{app:radiationtransport}.
 
In Fig.~\ref{fig:lc} we show the light curve for the uniform mix model, plotting separately the gammas from nuclear decays and the gammas from positron annihilation. We also show the prompt nuclear decay emission for comparison with the processed emission. This comparison shows that the ejecta become optically thin after approximately four years for this model. We also identify the isotope with the most significant contribution at each time and find that, similar to CCSNe, the light curve is dominated by $^{56}{\rm Ni}$ and $^{56}{\rm Co}$ decay at early times and is dominated by $^{44}{\rm Ti}$ at a few tens to a few hundred years. 
The left panel of Fig.~\ref{fig:lc} shows that the gamma-ray light curve summed over all photon energies is primarily composed of contributions from isotopes typically synthesized in standard neutrino-driven core-collapse supernovae. Several key isotopes, such as \( ^{57}\mathrm{Co} \), \( ^{44}\mathrm{Ti} \) (with its decay product \( ^{44}\mathrm{Sc} \)), and \( ^{60}\mathrm{Co} \), emit prominent gamma rays in distinct energy regions: \( ^{57}\mathrm{Co} \) and \( ^{44}\mathrm{Ti} \) dominate the signal with energy below 200\,keV, while \( ^{44}\mathrm{Sc} \) and \( ^{60}\mathrm{Co} \) dominate the energy above 1\,MeV.

The full-energy light curve can mask important contributions from isotopes whose gamma-ray emission lies between these energy regimes. To address this and also avoid the positron annihilation signal, we examine the light curve in the 550--750\,keV energy window and present the results in the right panel of Fig.~\ref{fig:lc}. The key difference in this narrower energy region is that, once the major contributors \( ^{56}\mathrm{Ni} \) and \( ^{56}\mathrm{Co} \) have decayed, the light curve becomes dominated by second-peak $r$-process isotopes such as \( ^{125}\mathrm{Sb} \), and \( ^{126}\mathrm{Sb} \).

%
%
%

\begin{figure}
\plottwo{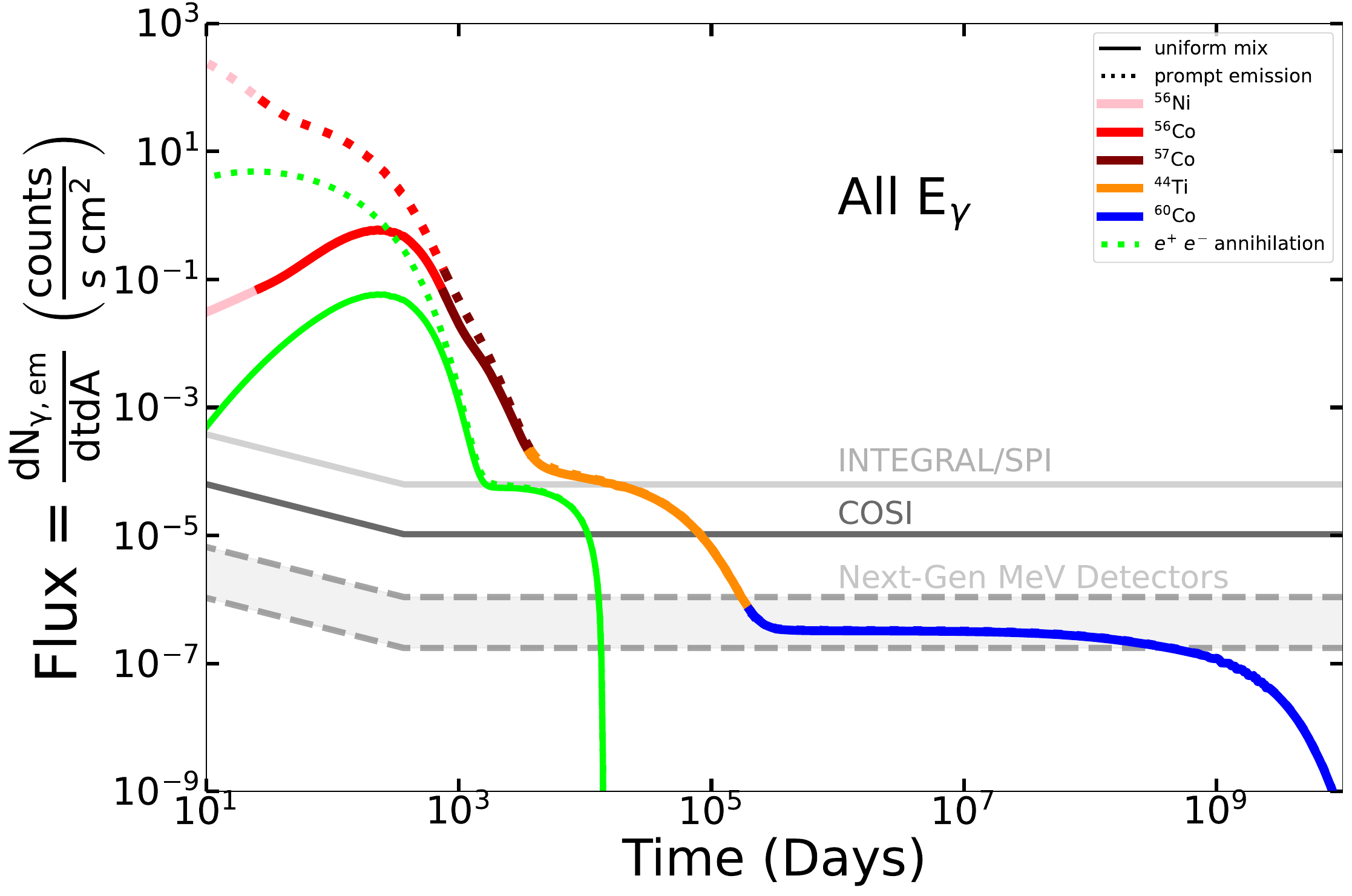}{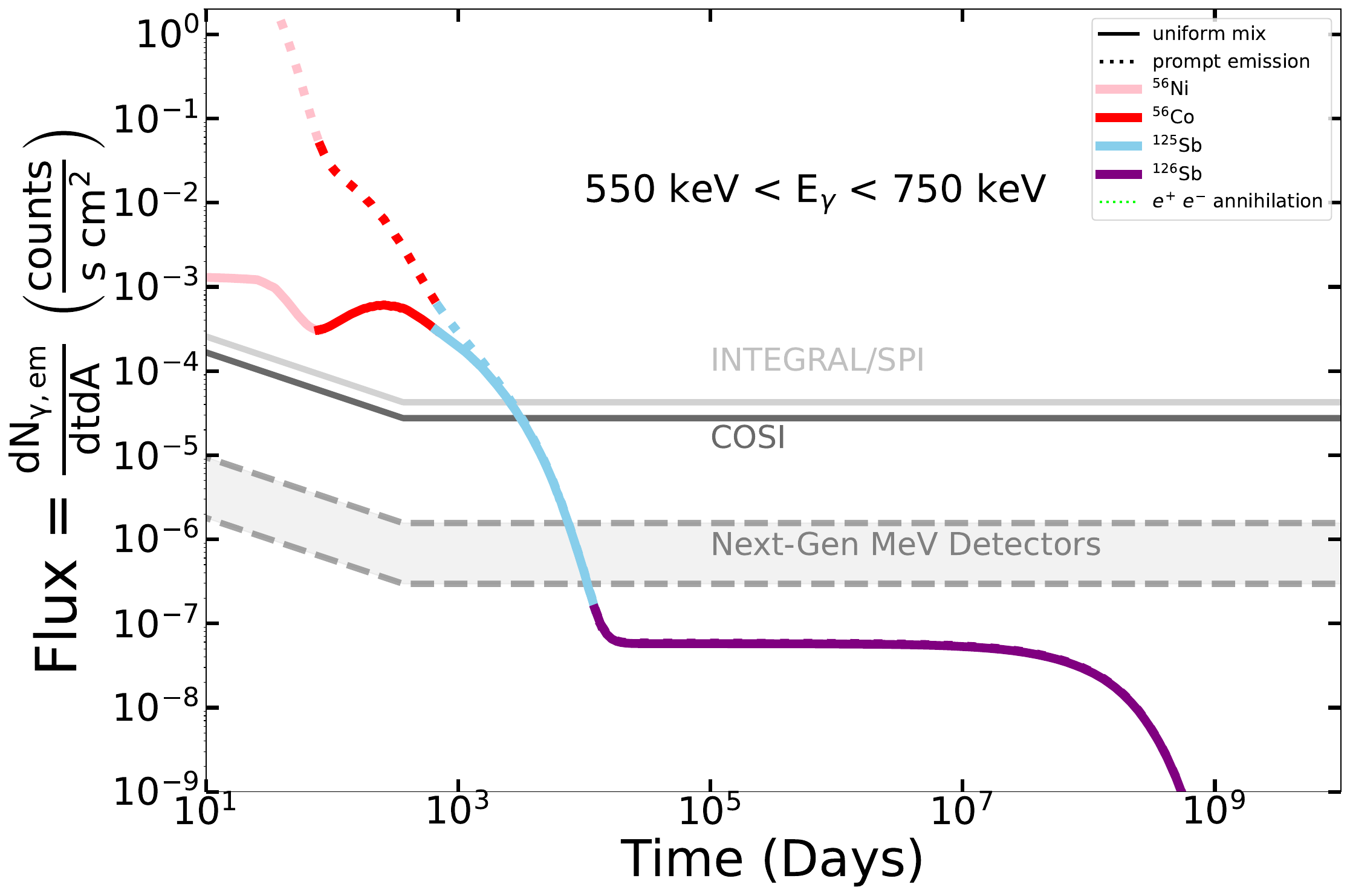}
\caption{The light curves of the beta-decay gamma-ray emission from the MR-SN shown in Fig. \ref{fig:snapshot}, over the whole energy range (left) and $550 \le E_{\gamma} \le
750$ keV (right),  assuming a distance of 10 kpc.
The curve is color-coded according to the decaying nuclear species with the largest contribution at a given time. The gray lines represent the sensitivity limits of different MeV instruments. For times $t$ less than 1 year post-explosion, the sensitivity curve is scaled with focusing time $=t$. For $t \geq 1$ year, we adopt a sensitivity that corresponds to one year. The green line represents an estimate of the flux due to electron-positron annihilation within the supernova remnant. Dotted lines indicate prompt (unabsorbed) emission, and solid lines indicate final emission (including absorption and Doppler broadening as in Sec. \ref{sec:doppler}) from the supernova.}
\label{fig:lc}
\end{figure}

\subsection{Comparing with observation}
\label{sec:instrument}
We compare the theoretically calculated gamma spectrum with the 3$\sigma$ sensitivity limits of MeV gamma-ray detectors.  We utilize published sensitivities for 
INTEGRAL/SPI \citep{2003A&A...411L..91R, 2020APh...114..107A}, COSI \citep{2022icrc.confE.652T, Tomsick:2023aue}, as well as next generation detectors such as AMEGO \citep{McEnery2019All} and GRAMS \citep{2020APh...114..107A}.  A summary of published continuum sensitivities and line sensitivities for these detectors, which assume a $3 \sigma$ significance, at various energies is given in Tables \ref{tab:contsens} and \ref{tab:linesens} in Appendix \ref{app:sens}.  

Continuum sensitivities scale approximately as $S_c \propto \left(\Delta \Omega / t \right)^{1/2} $,  where $\Delta \Omega$ is the angular resolution, and $t$ is the observing time.  Values of $S_c$ are typically quoted in units of photon energy per area per time and can be converted to counts per area per time through a factor of $E^2 / \Delta E $, where $E$ is the gamma energy and $\Delta E$ is the energy range over which photons are collected.  In Fig. \ref{fig:lc} we compare our integrated number flux of photons, i.e. in units of ${\rm photons} \,  {\rm cm}^{-2} \, {\rm s}^{-1}$, to the various instruments discussed above with the continuum sensitivities from Table \ref{tab:contsens} converted using the typical $\Delta E = E/2$.  
We use the continuum sensitivity at 1 MeV and 600 keV as the representative energy for the corresponding energy range in the left and right panel respectively. We take the observation time to be the same as the time post-explosion until it reaches one year, after which it remains at one year.  We use the sensitivities of AMEGO (representing a similar sensitivity level to eAstrogam and MeVGRO) and GRAMS to illustrate the detecting performance range of the \lq Next-Generation MeV missions'. 

In Section \ref{sec:remnant}, we present estimates of spectra from CCSN remnants.  In these cases, instead of comparing with continuum sensitivities,  we utilize narrow line sensitivity limits, which also roughly scale as $S_l \propto \left(\Delta \Omega / t \right)^{1/2} $ and are quoted in dimensions of counts per area per time.  In contrast to the continuum sensitivity limits, which assume a collection of photons over a substantial energy range $\Delta E$, line sensitivities are determined by considering only a narrow spread in energy (related to the energy resolution of the detector) and are specific for each line. Therefore, for each detector, we start with the published narrow-line sensitivities, $S_l$, tabulated in Table \ref{tab:linesens}, and rescale them with the observing time $t$.  When no published value is available for a specific energy channel, we extract the sensitivity data by digitizing published line sensitivity curves using WebPlotDigitizer\footnote{\url{https://automeris.io/}}, and interpolate the resulting data points as needed. When considering remnants, we adopt an observing time of one year.

%
%

\section{Known supernova remnants and signatures of neutron-rich nucleosynthesis}
\label{sec:remnant}

Both panels of Figure~\ref{fig:lc}  show 
gamma-ray emission persisting beyond \(10^5\) days (\(\approx\) 275 years) after the explosion. These long-lasting signals are primarily due to gammas emitted from the beta minus decay of \(^{126}\mathrm{Sb}\) and \(^{60}\mathrm{Co}\).  While these nuclei are not particularly long-lived, they are produced by long-lived parent nuclei.  The \(^{126}\mathrm{Sb}\) signal, which includes lines at 666\,keV and 695\,keV, arises from the decay of its long-lived parent \(^{126}\mathrm{Sn}\), which has a half-life of \(t_{1/2} \approx 1.98 \times 10^5\) years. Similarly \(^{60}\mathrm{Co}\) is the daughter product of the long-lived isotope \(^{60}\mathrm{Fe}\) that has a half-life of \(t_{1/2} \approx 2.61 \times 10^6\) years. The beta minus decay of \(^{60}\mathrm{Co}\) produces gamma-ray lines at 1173\,keV and 1332\,keV.  Figure~\ref{fig:lc} compares the gamma signal with detector sensitivities for a supernova remnant at 10 kpc, although several supernova remnants are closer than this.

%
%
\begin{table}[]
\centering
\begin{tabular}{|c|l|l|l|l|l|l|}
\hline
Label & Remnant name & Age (yrs) & Distance (kpc) & Reference & Adopted age (yrs) & Adopted distance (kpc) \\
\hline
\textcolor{blue}{\textbullet}& Cassiopeia A     & $\sim$340   & 3.3  & 1, 2 & 340 & 3.3 \\ \hline
 \textcolor{blue}{\textbullet}& Crab Nebula      & 971   & 1.9  & 1 & 971 & 1.9 \\ \hline
 \textcolor{blue}{\textbullet}& IC 443           & 3000-30000 & 1.5-2.5   & 3, 4, 5 & 30000 & 1.5 \\ \hline
 \textcolor{blue}{\textbullet}& Vela Junior      & 2400-5100  & 0.5-1   & 6 & 2400 & 0.7 \\ \hline
 \textcolor{blue}{\textbullet}& Vela SNR         & $\sim$11400 & 0.25-0.3  & 1, 7 & 11400 & 0.25 \\ \hline
 \textcolor{blue}{\textbullet}& SN1987A          & 37    & 50    & 2, 8 & 37 & 60 \\ \hline
a & Kes 75           & $\sim$480   & 5.6  & 1, 9 & 480 & 5.6 \\ \hline
b & G350.1-0.3       & $\sim$600   & 4.5  & 10 & 600 & 4.5 \\ \hline
c & G330.2+1.0       & $\sim$1000  & 5  & 11 & 1000 & 5 \\ \hline
d & G320.4-1.2       & $\sim$1700  & 3-5.2  & 1, 12 & 1700 & 3 \\ \hline
e & RCW103           & $\sim$2000  & 3  & 1, 13 & 2000 & 3 \\ \hline
f & G11.2-0.3        & 1400-2400  & 4.7  & 1, 14 & 2400 & 4.7 \\ \hline
g & W49              & 1000-4000  & $\sim$10  & 1, 15 & 1000 & 10 \\ \hline
\end{tabular}
\caption{
Galactic and nearby extragalactic Type II supernova remnants we include in Fig. \ref{fig:126sb_observe_0} to calculate the potential observability of \(^{126}\)Sb and \(^{60}\)Co in MeV lines. Reference: 1-\cite{2025Green},2-\cite{2020A&A...638A..83W}, 3-\cite{2023A&A...680A..83C}, 4-\cite{2025arXiv250112613A}, 5-\cite{2017MNRAS.472...51A}, 6-\cite{2015ApJ...798...82A}, 7-\cite{1970ApJ...159L..35R}, 8-\cite{2019Natur.567..200P}, 9-\cite{2018ApJ...856..133R}, 10-\cite{2020ApJ...905L..19B}, 11-\cite{2018ApJ...868L..21B}, 12-\cite{1982ApJ...262L..31M}, 13-\cite{2023ApJ...958...30S}, 14-\cite{2016ApJ...819..160B}, 15-\cite{2014ApJ...793...95Z}.  
\label{tab:remnant}}
\end{table}

\begin{figure}
    \centering
    \includegraphics[width=0.7\linewidth]{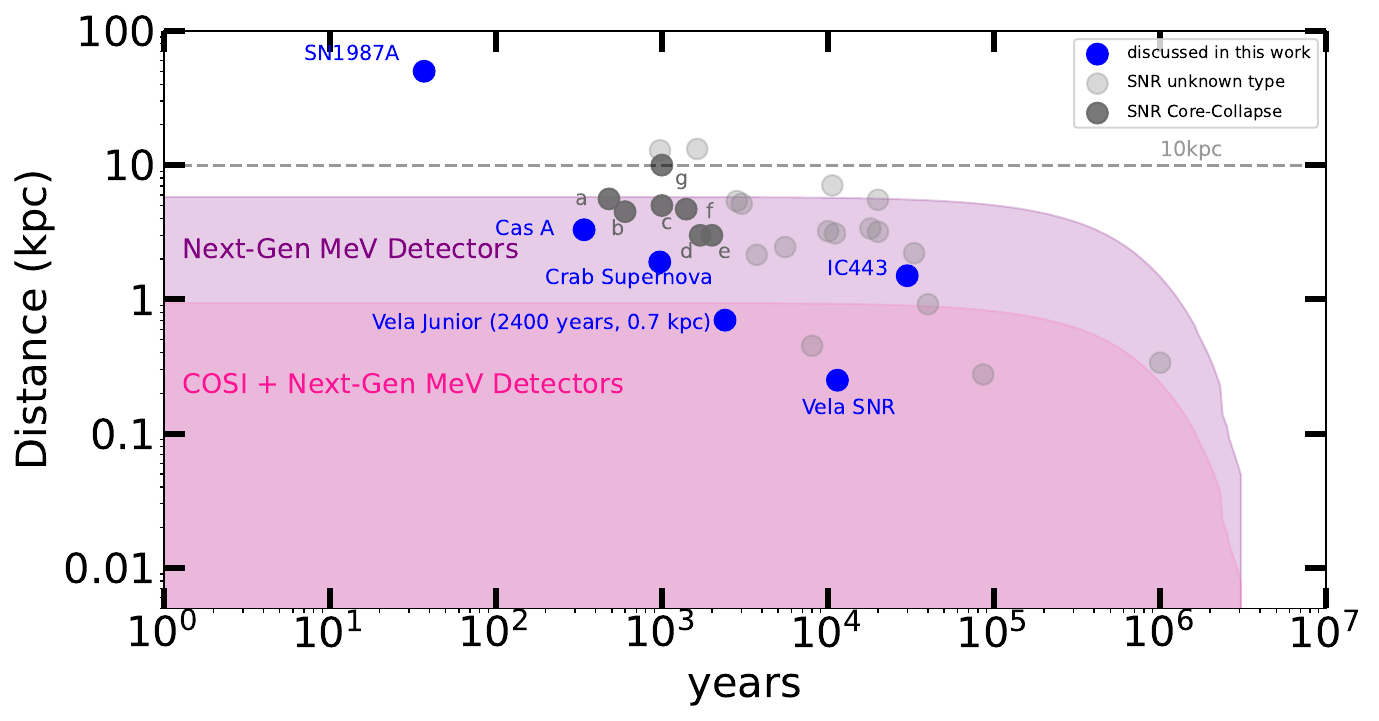}
    
    \caption{Observability of $^{126}$Sb 666 keV line with COSI (light and dark pink region) and next-generation telescopes (light pink region) if a supernova remnant originated from an MR-SN described in Fig. \ref{fig:snapshot}.  
    The x-axis denotes the age and the y-axis denotes the distance of the remnant.  Labels for supernova remnants are as follows: a: Kes75, b: G350.1-0.3, c: G330.2+1.0, d: G320.4-1.2, e: RCW103, f: G11.2-0.3, g: W49. 
    }
    \label{fig:126sb_observe_0}
\end{figure}

Ages and distances for thirteen CCSN remnants are listed in Table \ref{tab:remnant}.  Where available, uncertainties are also listed, although we note that even when not listed explicitly, there are substantial uncertainties in the age and distance estimates of supernova remnants. For this work, we adopt the age and distance listed in the last two columns of the table, and we plot these as blue and dark gray dots in Fig.~\ref{fig:126sb_observe_0} and Fig.~\ref{fig:60co_observe_2}. In addition, we show supernova remnants of unknown type as light gray dots, using age and distance estimates from \cite{2025Green}. Later in Section~\ref{sec:remnants}, we utilize the remnants labeled in blue to perform a detailed spectral analysis, investigating late-time gamma-ray signatures.

%
%

\subsection{$^{126}{Sn}$ $(^{126}{Sb})$ in remnants}
Our model yields \(2.5 \times 10^{-4} M_\odot\) of \(^{126}\)Sb.
This value is lower than the yields from \citet{2017ApJ...836L..21N} but higher than the yields in \cite{2015ApJ...810..109N} and higher than a subset of the models in \cite{2021MNRAS.501.5733R}. 
We use this yield and  line sensitivities from COSI 
and GRAMS to determine the maximum distance at which a given line could be observable. 
For the  666 keV and 695 keV lines produced by the decay of $^{126}{\rm Sb}$, we adopt a line sensitivity of $S_l =  4 \times 10^{-6} {\rm ph}/ ({\rm cm}^2 {\rm s})$ for COSI and $S_l = 7 \times 10^{-8} {\rm ph}/ ({\rm cm}^2 {\rm s})$ for the next-generation detector, GRAMS. 
For COSI, this corresponds to an observational range of roughly 1 kpc; for next-generation detectors, this corresponds to about 6 kpc.  This range is shown in Fig.~\ref{fig:126sb_observe_0} as a dark pink region for COSI, while the light pink region shows the additional range available with next-generation MeV detectors. The nearly flat plateau observed in Fig. \ref{fig:126sb_observe_0} is due to the long-lived nature of $^{126}{\rm Sn}$.  

As shown in Fig. \ref{fig:126sb_observe_0}, about 11 confirmed CCSN remnants are close enough to be above detector sensitivities for  $^{126}{\rm Sn}$ ($^{126}{\rm Sb}$). Our estimates suggest that were $^{126}$Sb to be produced in Vela Junior or Vela SNR, the associated gamma-ray flux would be above detector sensitivity of COSI. Additionally,  
the Crab SNR, Jellyfish SNR (IC441), Cassiopeia A, Kes75, G350.1-0.3, G330.2+1.0, G320.4-1.2, RCW103, and G11.2-0.3 are all within the sensitivity range of next-generation MeV gamma-ray missions. Such a detection would constitute the discovery that the synthesis of the second peak $r$-process element $^{126}$Sn occurred in a nearby supernova explosion. At the same time, a non-detection would place a limit on the amount of $r$-process that could have been produced in the corresponding event.  If more information becomes available about the supernova remnants of unknown type that fall within the observational range, it would present additional opportunities.

\subsection{$^{60}{Fe}$ $(^{60}{Co}$) in remnants}

%
%
\begin{figure}
   \centering
   \includegraphics[width=0.7\linewidth]{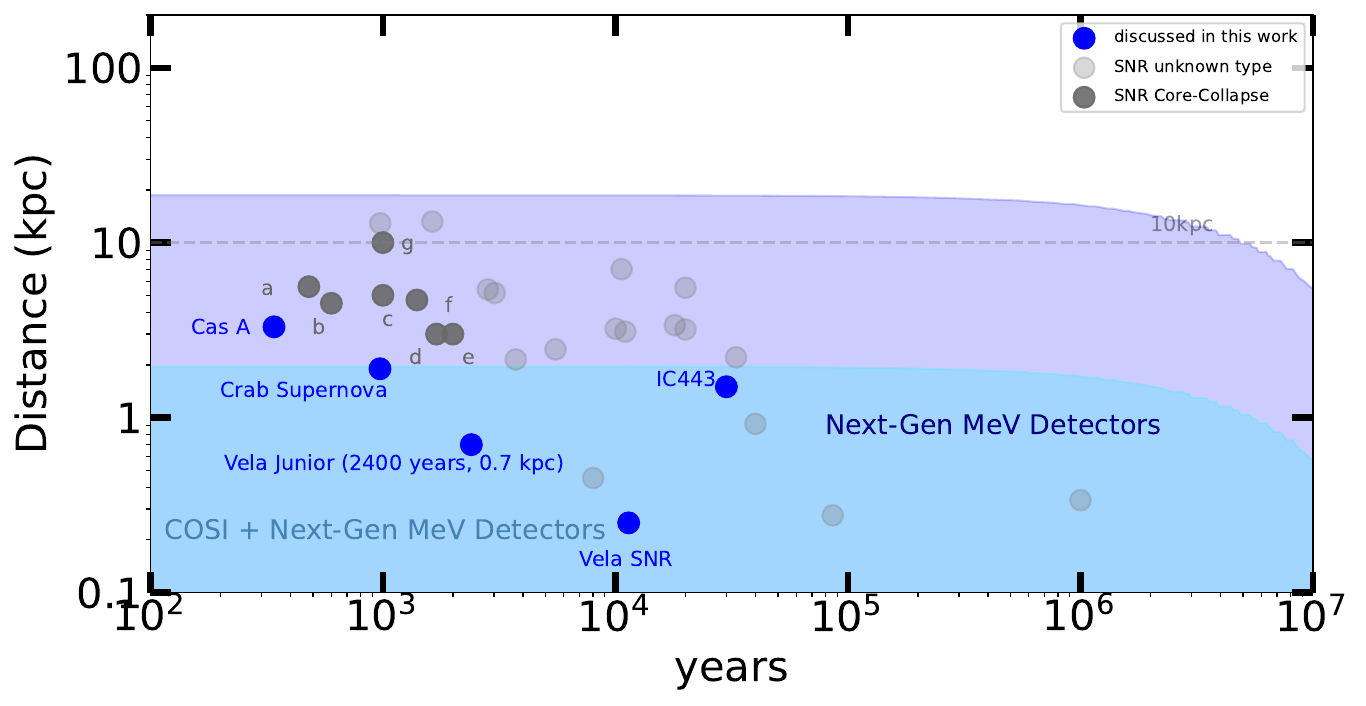} \\
   \includegraphics[width=0.7\textwidth]{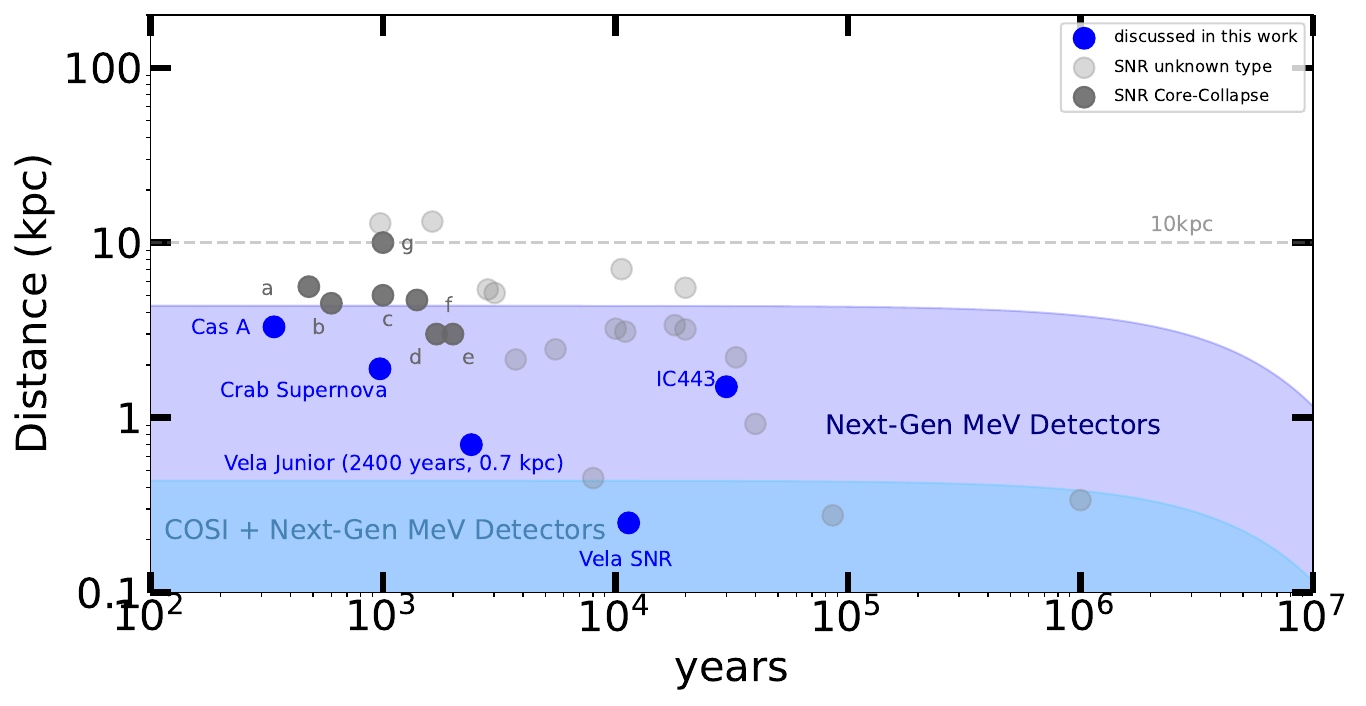}
   \caption{Top panel: Same as Fig. \ref{fig:126sb_observe_0} but for the observability of $^{60}$Co 1332 keV line using the $^{60}{\rm Fe}$ yield from the MN-SN model. Bottom panel: same as the left panel, but for CCSN.  The $^{60}$Fe yield is taken from \cite{2018Limongi} as $3.7 \times 10^{-5}$ M$_{\odot}$
   }
   \label{fig:60co_observe_2}
\end{figure}

The 1173 and 1332 keV lines from the decay of $^{60}{\rm Co}$ have been detected in the interstellar medium \citep{2007A&A...469.1005W,2011ApJ...739...29B} but have not yet been detected in supernova remnants.
Such an observation would be the first direct evidence of in situ $^{60}{\rm Fe}$ production.
It would be complementary to the theoretical yields of $^{60}{\rm Fe}$ in neutrino/advection-driven CCSN as well as indirect observational evidence, such as the ratio of $^{26}{\rm Al}$ to $^{60}{\rm Fe}$ in the interstellar medium \citep{Jones:2019des,2007A&A...469.1005W,2011ApJ...739...29B}, the deposition of $^{60}{\rm Fe}$ on the sea floor \citep{2016Wallner}, lunar samples \citep{2014LPI....45.1778F,2012LPI....43.1279F}, and Antarctic snow \citep{2019PhRvL.123g2701K}. The latter three types of measurements suggest a relatively recent and nearby CCSN event \citep{2004PhRvL..93q1103K,2015ApJ...800...71F,2019MNRAS.486.2910W,2021ApJ...923..219W,2023ApJ...947...58E}. 

For CCSNe, the production mechanism for $^{60}{\rm Fe}$ is a combination of pre-supernova hydrostatic burning and explosive burning due to the supernova shock passage \citep{2006NewAR..50..474L,Jones:2019des,2020andrews,2024Burrows}. 
In Table \ref{tab:60Fe_yields} in Appendix \ref{sec:60Feyields}, we list several published model yields. We note that the yields depend critically on reaction rates \citep{2024NatCo..15.9608S}. We adopt the zero rotation, [Fe/H]~=~0, 25 M$_{\odot}$ model from \cite{2018Limongi} for the CCSN $^{60}{\rm Fe}$ yield, which is ${3.7 \times 10^{-5}}$~M$_{\odot}$.  
We use this to estimate the observability range of the 1332 keV line of CCSN $^{60}$Co using detector sensitivities of ${S_l = 3 \times 10^{-6} {\rm ph}/ ({\rm cm}^2 {\rm s})}$ for COSI and ${S_l = 3 \times 10^{-8} {\rm ph}/ ({\rm cm}^2 {\rm s})}$ for next-generation detectors (GRAMS). 
As can be seen from the bottom panel of Fig.~\ref{fig:60co_observe_2}, we do not expect to be able to observe $^{60}$Co beyond about 4-5 kpc if a standard neutrino/advection-driven CCSN produced the supernova remnant. Only the remnant, Vela, is within the COSI detection region. However, Vela is an extended source with a large angular size, so it deserves further study before conclusions can be drawn about its expected $^{60}{\rm Co}$ signal.
Six additional remnants, Cassiopeia A, Jellyfish, Crab, Vela Junior, G320.4- 1.2, and RCW 103, are within the next-generation telescope range. 

MR-SNe models show production of orders of magnitude more $^{60}{\rm Fe}$ due to the neutron rich conditions in the jet \citep{2023MNRAS.518.1557R}, ranging from $2\times10^{-3} \, {\rm M}_\odot$ to $5.7\times10^{-3} \, {\rm M}_\odot$ as shown in Table \ref{tab:60Fe_yields}.
Even in MR-SN models where the main $r$-process is not significantly produced, the yield of $^{60}{\rm Fe}$ is considerable. 
In our model the $^{60}$Fe yield is $8.3\times10^{-3}$ M$_{\odot}$ and is consistent with the yield of $^{60}{\rm Fe}$  in other MR-SN studies \citep{2015ApJ...810..109N,2017ApJ...836L..21N}. 
Using this yield, we show the distance range where the flux from the 1332 keV line is above the sensitivity limits in the top panel of Fig. \ref{fig:60co_observe_2}.  This is approximately 2 kpc for COSI and about 18 kpc for next-generation detectors. 
We see that a total of twelve CCSN remnants (all the Galactic remnants listed in Table \ref{tab:remnant}) could have $^{60}$Co 1332 keV line above sensitivity limits if an MR-SN produced that remnant.
 
Comparison of the top and bottom panels of Fig. \ref{fig:60co_observe_2} highlights the difference in the $^{60}{\rm Fe}$ observability range for the two types of supernova. If $^{60}$Co were detected in supernova remnants beyond 0.3~kpc by COSI, or beyond 4-5 kpc by next-generation gamma-ray instruments, such a positive detection would favor an MR-SN origin. 
However, the situation is more ambiguous for remnants located within approximately 4 kpc, such as Cassiopeia A.  
An estimate of the total mass of synthesized $^{60}$Co or the companion detection of $^{126}$Sb (see the following section) would be required to discriminate between explosion mechanisms. 
In contrast, a non-detection within any of the predicted observation ranges would constrain supernova modeling and/or the proposed production mechanism of $^{60}$Fe.

\subsection{Implications of a combined analysis of $^{60}{Fe}$ and $^{126}{Sn}$ in remnants}

We consider what can be learned from a combined analysis of constraints on observations of $^{126}{\rm Sb}$ and $^{60}{\rm Co}$ abundances.  
From Figs.~\ref{fig:126sb_observe_0} and \ref{fig:60co_observe_2}, we see that the $^{60}{\rm Co}$ signal is stronger than that of $^{126}$Sb, due primarily to the larger predicted yield of $^{60}{\rm Fe}$ as compared with $^{126}{\rm Sn}$.  
For a sufficiently nearby remnant, both lines should accompany each other if the remnant is the result of an $r$-process-producing MR-SN.  
However, the situation is less clear if only $^{60}{\rm Co}$ is observed.  
The remnant could be due to either an MR-SN or an advection/neutrino driven CCSN, although the magnitude of an observationally derived abundance limit would hint at one scenario or the other.  A substantial abundance of $^{60}{\rm Co}$ without $^{126}{\rm Sb}$ would hint at an MR-SN that has a nucleosynthesis pattern that does not make it as far as the second peak of the $r$-process. A more modest measured abundance of $^{60}{\rm Co}$, would suggest a CCSN origin and since the range of $^{60}{\rm Fe}$ abundance predictions is wide, it would provide a diagnostic for CCSN models.  Although unlikely, if only $^{126}$Sb is detected in these remnants, it would inspire us to revisit all types of supernova modeling and to explore more exotic $r$-process-producing supernova scenarios.

\subsection{Cassiopeia A, Crab Supernova, IC443 and Vela Junior}
\label{sec:remnants}

\begin{figure}[htbp]
    \centering
    \begin{subfigure}
        \centering
        \includegraphics[width=1\linewidth]{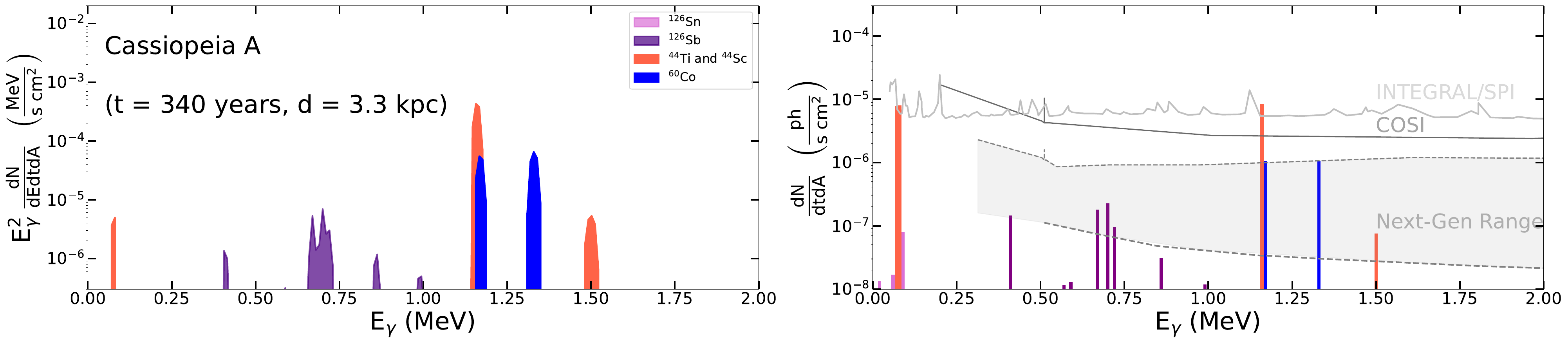}
    \end{subfigure}
    
    \begin{subfigure}  
        \centering
        \includegraphics[width=1\linewidth]{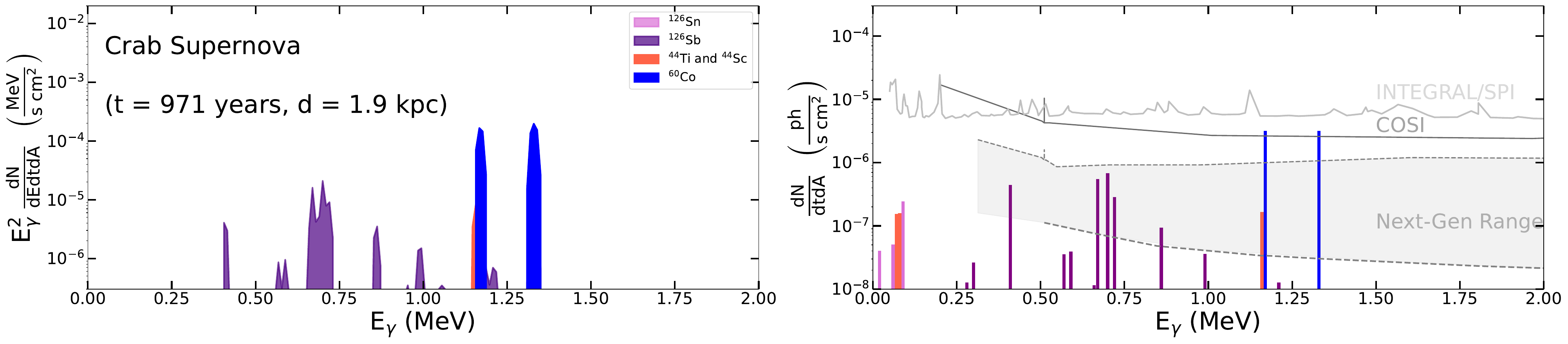}
    \end{subfigure}
    
    \begin{subfigure}  
        \centering
        \includegraphics[width=1\linewidth]{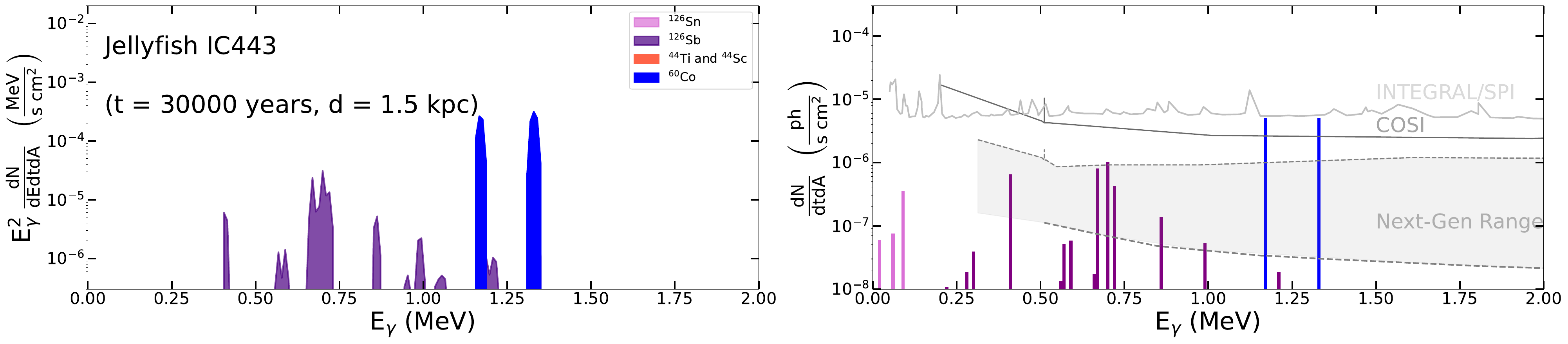}
    \end{subfigure}
    
    \begin{subfigure}
        \centering
        \includegraphics[width=1\linewidth]{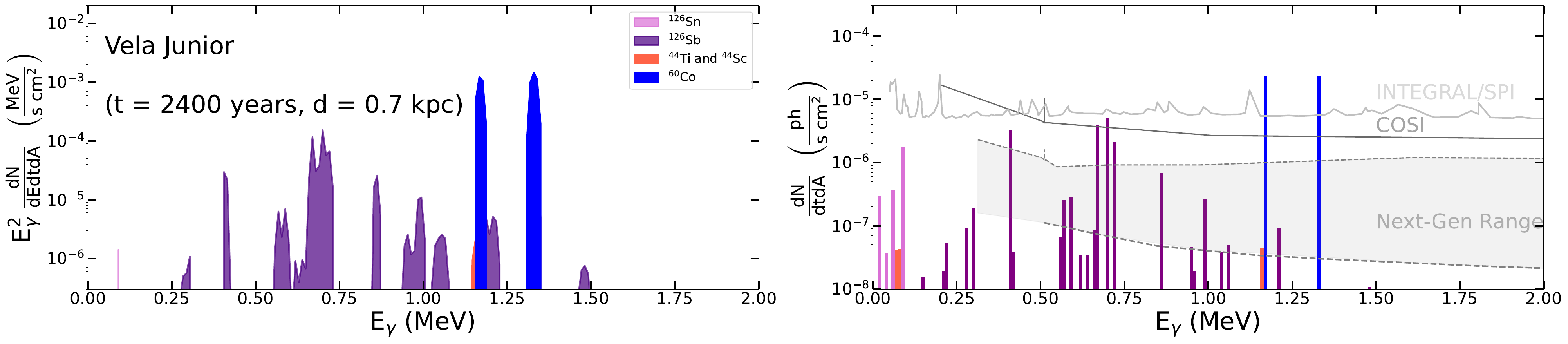}
    \end{subfigure}
    
    \caption{
    Gamma-ray spectra of four known supernova remnants, assuming they are remnants from the MR-SN event shown in Fig. \ref{fig:snapshot}. The left panel shows the Doppler-broadened spectra due to the expansion of the remnant. The right panel shows the total number of prompt photons per area per time for a direct comparison with the line sensitivity limits. Detector sensitivities are colored gray with the same line style as in the Fig. \ref{fig:lc} 
    and have been scaled to 1 year. These late-time remnants feature \(^{126}\)Sb as the dominant $r$-process signal. Gamma emission from \(^{60}\)Co (\(^{60}\)Fe) are prominent between 1 MeV to 1.5 MeV.}
    \label{fig:combined}
\end{figure}

In Fig. \ref{fig:combined}, we show predicted spectra for four supernova remnants, all of which fall within the range of next generation instruments for $^{126}$Sb ($^{126}$Sn) flux.  In the left column, we show Doppler-broadened spectra plotted in units of MeV/(s cm$^{2}$), using the standard relationship $E_\gamma^2  dN/(dE_\gamma dAdt)$.
The late-time spectrum is sparse and consists of decays from only a few prominent, long-lasting radioactive isotopes. Thus, we want to compare features with narrow line instrument sensitivities. To make a direct comparison, we plot the total photons at their emitted energy in units of photons/s/cm$^2$ in the right column along with sensitivities from Table \ref{tab:linesens}\footnote{GRAMS does operate below 511 keV, however, the line sensitivity information below 511 keV is not given in \cite{2020APh...114..107A}.}.  Looking at the left column of Fig. \ref{fig:combined}, we see that in all four supernova remnants, Cassiopeia A, Crab, Jellyfish and Vela Junior, the $^{126}{\rm Sb}$ lines are above next generation detector sensitivity, and in the case of Vela Junior, two of the lines are just slightly above COSI sensitivity. In Vela Junior, the $^{60}{\rm Fe}$ prediction for MR-SNe suggests a flux from $^{60}{\rm Co}$ decay, which is above both COSI and SPI sensitivities, and COSI sensitivity for Crab and Jellyfish.

In addition to $^{126}{\rm Sb}$ and $^{60}{\rm Co}$, in Fig. \ref{fig:combined}, we also see in orange the 68 keV and 78 keV lines which accompany \(^{44}\)Ti decay and the 1157 keV line which accompanies the subsequent beta plus decay of \(^{44}\)Sc. For Cassiopeia A, INTEGRAL/SPI \citep{2015A&A...579A.124S,2020A&A...638A..83W}, NuSTAR \citep{2014Grefenstette,2017ApJ...834...19G}, INTEGRAL IBIS/ISGRI \citep{2006Renaud}, CGRO/COMPTEL \citep{1994A&A...284L...1I}, Bepposax \citep{2001ApJ...560L..79V}  have reported on the 68 keV, 79 keV, and 1157 keV lines of \(^{44}\)Ti and \(^{44}\)Sc; see Table \ref{tab:44and56yields_comparison} and \cite{2016MNRAS.458.3411T} for a summary.  Estimates of the mass of  \(^{44}\)Ti in Cassiopeia A from observation are known to be on the upper end of theoretical predictions; see columns 4 and 5 of Table \ref{tab:44and56yields_comparison}.  Since in our model, the yield of  \(^{44}\)Ti is similar to the theoretical estimates in CCSNe, we do not anticipate that the \({} ^{44}\)Ti will be a helpful diagnostic of the explosion mechanism without a more detailed study.

%
%

%
%
%

\begin{table}[h!]
\centering
\begin{tabular}{|c|c|c|c|c|}
\hline
                     & This Study & SN1987A                                                   & Cassiopeia A & Other theoretical studies       \\ \hline
$^{44}$Ti ($10^{-4}$M$_{\odot}$) & 0.42 & \begin{tabular}[c]{@{}c@{}}$0.55\pm0.17^{a}$  \\ $1.5\pm0.3^{b}$  \\ $3.1 \pm 0.8^{c}$  \end{tabular} & \begin{tabular}[c]{@{}c@{}}$2.6\pm0.6^{e}$  \\ $1.5\pm0.2^{d}$  \\ $2.4 \pm 0.9^{f}$  \end{tabular} & \begin{tabular}[c]{@{}l@{}}1$^{i}$\\ 0.2 to 0.5$^{j}$\\ 1 to 2$^{k}$\end{tabular}            \\ \hline
$^{56}$Ni ($10^{-2}$M$_{\odot}$) & 8 & 7$^{h}$                                                & 7 to 15$^{g}$ & \begin{tabular}[c]{@{}l@{}}12$^{i}$\\ 7$^{j}$\\ 10 to 15$^{k}$\end{tabular}\\ \hline
\end{tabular}
\caption{Comparison of $^{44}$Ti and $^{56}$Ni yields among different studies. a: \cite{2014Seitenzahl}, b: \cite{2015Boggs}, c: \cite{2012Grebenev}, d: \cite{2017ApJ...834...19G}, e: \cite{2020A&A...638A..83W}, f: \cite{2015Siegert}, g: \cite{2008Krause}, h: \cite{1989Bussard}, \cite{1988Leising}, \cite{2015Dwek}, \cite{2005Vink}, i: \cite{2016Sukhbold} 25.2$M_\odot$ with W18 engine, j: \cite{2018Limongi} 25$M_\odot$ initial mass with various [Fe/H] and rotational speed, k: \cite{2024Wang} $M_{\text{ZAMS}}$ between 20-60$M_\odot$.}
\label{tab:44and56yields_comparison}
\end{table}

 For relatively young supernova remnants such as Vela Junior and the Crab, the 1157 keV line from $^{44}$Sc may fall within the detectable range of next-generation gamma-ray instruments. However,  Doppler broadening could make it difficult to distinguish this line from the nearby 1173 keV line of $^{60}$Co. If the intensity of the $^{44}$Sc line is relatively low as is the case for Crab, it may be obscured by the $^{60}$Co feature; conversely, if the $^{60}$Co signal is weaker, as it is for Cassiopeia A, the $^{60}$Co line may be challenging to detect.

The beta plus decay of $^{44}{\rm Sc}$ has additional implications for MeV gamma detection, since the emitted positron may slow down and annihilate before leaving the remnant. However, when we estimate this annihilation signal using the method described in Sec. \ref{sec:annihilation}, we find that for these four remnants the annihilation contribution is negligible, and most positrons produced by the decay of $^{44}{\rm Sc}$  escape before annihilating.  

\subsection{SN 1987A} 
\begin{figure}
    \centering
    \includegraphics[width=1\linewidth]{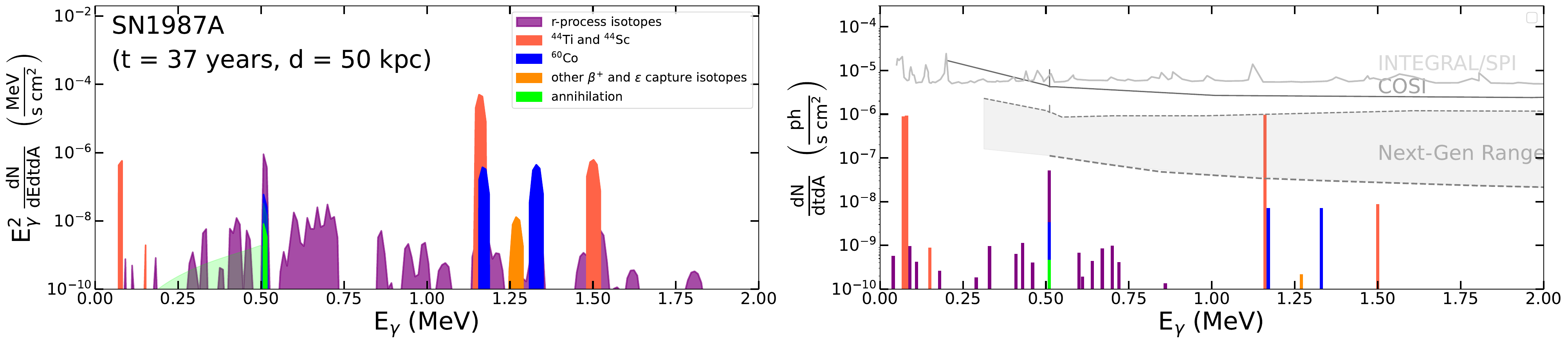}
    \caption{Similar to the Fig. \ref{fig:combined} but for SN1987A. The 1.157 MeV \(^{44}\)Sc line previously reported by INTEGRAL in SN1987A \citep{2012Grebenev} is now weaker due to the currently older age of the remnant. 
    The 68 keV and 78 keV line from \(^{44}\)Ti have been detected by NuSTAR \citep{2015Boggs}. The \(^{44}\)Sc and \(^{44}\)Ti lines are also expected in neutrino/advection driven CCSN.}
    \label{fig:sn1987a}
\end{figure}

For completeness, we show the predicted spectrum of SN1987A, assuming the event was an MR-SN, in Fig. \ref{fig:sn1987a}. The 68 keV and 78 keV lines from $^{44}{\rm Ti}$ decay have been detected by IBIS and NuSTAR \citep{2012Grebenev,2015Boggs} which   
have better sensitivity in the tens of keV region than the MeV instruments that we compare with in this work. As SN 1987A resides far away at 50 kpc, we see in the figure that heavy isotopes from the $r$-process and $^{60}{\rm Co}$ are not within detection range for future MeV detectors. This is the only remnant for which the annihilation signal appears in the plot, but it is below the detectable limits.

%
%

%
%
\begin{figure}
    \centering
    \includegraphics[width=1\linewidth]{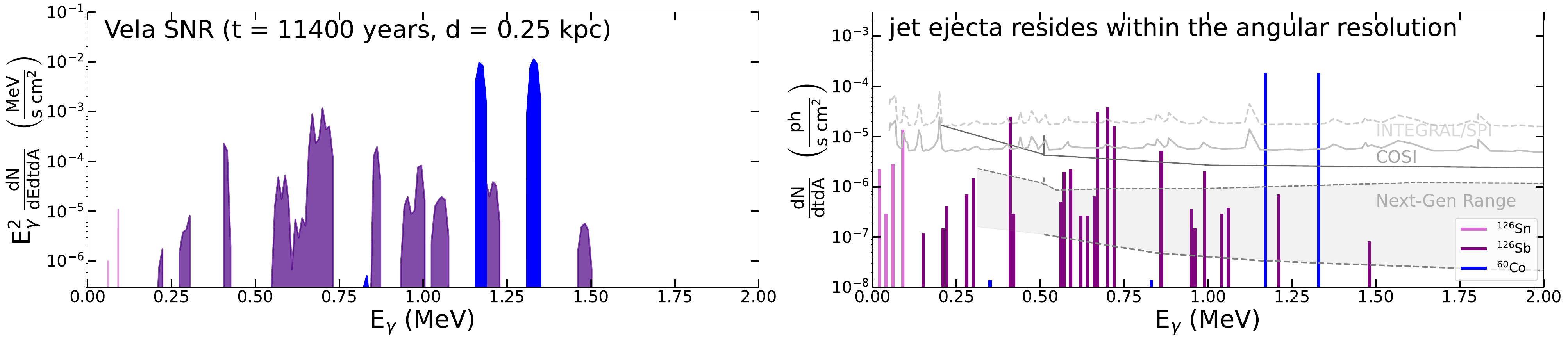}
    \caption{Similar to the Fig. \ref{fig:combined} but for Vela SNR.  For Vela,  we assume that the jet material is not mixed uniformly, so the jet ejecta resides within the angular resolution of the detector.  To estimate the reduction in sensitivity in the case where the ejecta are uniformly mixed, we scale the INTEGRAL/SPI line (dashed gray) by the ratio of the angular resolution of the instrument to the angular extent of Vela. 
    }
    \label{fig:vela-snr-scaled}
\end{figure}

\subsection{Vela SNR}

Vela SNR is located just 200~pc away, making the predicted gamma-ray lines of $^{126}{\rm Sb}$ and $^{60}{\rm Co}$ in our model particularly strong. However, Vela SNR is spatially extended and diffuse, spanning an angular size of approximately \(8^\circ\) on the sky \citep{2023A&A...676A..68M}, which is significantly larger than the angular resolution of most instruments. Fig. \ref{fig:vela-snr-scaled} therefore illustrates the optimistic scenario in which the remnant distribution is not homogeneous, so jet ejecta is confined within the angular resolution of the detector. 
According to the SPI observer's manual\footnote{\url{https://integral.esac.esa.int/AO14/AO14_SPI_ObsMan.pdf}}, the sensitivity roughly scales with the ratio of the instrument’s angular resolution to the source angular size. 
The SPI instrument has a full width at half maximum (FWHM) of \(\sim 2.5^\circ\) \citep{2003A&A...411L..91R}, COSI up to \(\sim 4^\circ\) at 511 keV \citep{Tomsick:2023aue}, and next-generation instruments offer resolutions around \(\sim 2^\circ\)--\(3^\circ\) FWHM \citep{McEnery2019All,2020APh...114..107A}. This disparity between source size and instrument resolution can reduce the sensitivity if the ejecta is uniformly distributed across the remnant.  This scaling effect is illustrated in Fig. \ref{fig:vela-snr-scaled} as the dashed gray line, representing the adjusted SPI 3$\sigma$ sensitivity. In fact, \citet{2007A&A...469.1005W} reported an  upper limit for \(^{60}\)Fe gamma-ray flux of  \(1.1 \times 10^{-5}~\text{ph cm}^{-2}\text{ s}^{-1}\) ($2 \sigma$) from SPI in the Vela region. This value is about an order of magnitude lower than our estimate, making an MR-SN less likely as an explanation for the Vela SNR. Specifically, in our model, we predict that the two strong $^{60}{\rm Co}$ lines would be above this upper limit.  However, because of its importance, the possibility of detecting $^{60}{\rm Fe}$ in Vela deserves further study, as do the 414 keV, 666 keV, and 695/697 keV lines of $^{126}{\rm Sb}$.

%
%
%

\section{Future Galactic Supernovae and signals of neutron-richness}
\label{sec:earlystage}
We now turn to the early-stage emission from a 
MR-SNe event, should it occur within the Galaxy. 
As shown in the left panel of Fig.~\ref{fig:lc}, the early light curve is dominated by emission from the decay of $^{56}$Ni which has a half-life of $t_{1/2} \approx 6$ days and $^{56} {\rm Co}$ which has a half-life of $t_{1/2} \approx 77$ days. In our model, the total amount of \(^{56}\)Ni ejected to be about $0.08 M_\odot$, which aligns well both with other theoretical predictions and with estimates from observations of Cassiopeia A and SN 1987, see Table \ref{tab:44and56yields_comparison}.  The observational values in this table are based on modeling of the light curve for SN 1987A (\cite{1989Bussard}, \cite{1988Leising}, \cite{2015Dwek}, \cite{2005Vink})  and modeling optical observations of Cassiopeia A \citep{2008Krause}. 

The strongest gamma lines from  $^{56}$Ni and $^{56} {\rm Co}$ have intensities of P(158 keV)=98\% and  P(811 keV)=86\%, for $^{56}$Ni, and  P(846 keV) = 99\%  and P(1238 keV) = 66\% for $^{56}$Co. These decays also dominate neutrino/advection driven CCSN events.  The 846 keV and 1238 keV lines from  \(^{56}\)Co decay were detected by the Gamma-Ray Spectrometer (GRS) onboard the Solar Maximum Mission (SMM) satellite \citep{1988Natur.331..416M} when SN1987A shortly after explosion \citep{1988Natur.331..416M}.  The strength of these lines corresponds to a few percent of the \(^{56}\)Ni mass derived from light curve modeling.

By 1000 days our model suggests that these lines have faded considerably, and the light curve becomes dominated by $^{57}$Co, which has a half-life of $t_{1/2} \approx 271$ days.  In our model, 0.0073 $M_{\odot}$ of \(^{57}\)Co is produced in total, with about one third of the \(^{57}\)Co being created in the jet and two thirds being produced during the passage of the shock.   \cite{Kurfess1992} reported the detection of the 122 keV \(^{57}\)Co line from SN1987A with the Oriented Scintillation Spectrometer Experiment (OSSE) aboard the Compton Gamma-Ray Observatory. The corresponding mass of \(^{57}\)Co  associated with two observations is between $0.002 M_{\odot}$ to  $0.004 M_{\odot}$.  The best fit of the BOVIR light curve by \cite{2014Seitenzahl} suggests that the \(^{57}\)Ni mass is around $ 0.004 \, M_{\odot}$, in agreement with the larger number reported in \cite{Kurfess1992} and is similar to our component of \(^{57}\) Co originating from explosive burning. 

Past 1000 days, the isotopes with the largest contributions are longer-lived isotopes such as $^{44}$Ti which has a half-life of 60 days and $^{60}$Co, which is seeded by the long-lived parent $^{60}$Fe. However, as discussed in Section~\ref{sec:doppler}, the color coding of the light curve in the left panel of Fig. \ref{fig:lc} reflects the isotopes with the largest contribution at each epoch 
but does not present 
all detectable features. For instance, the right panel of Fig.~\ref{fig:lc} shows that in the 550-750 keV energy range, gamma emission from $^{125}$Sb which has a half-life of 2.76 years  dominates at 1000 days.
Examining the full spectrum at all times is necessary to identify promising isotopes and decay lines.

\begin{figure}
    \centering
    \includegraphics[width=1\linewidth]{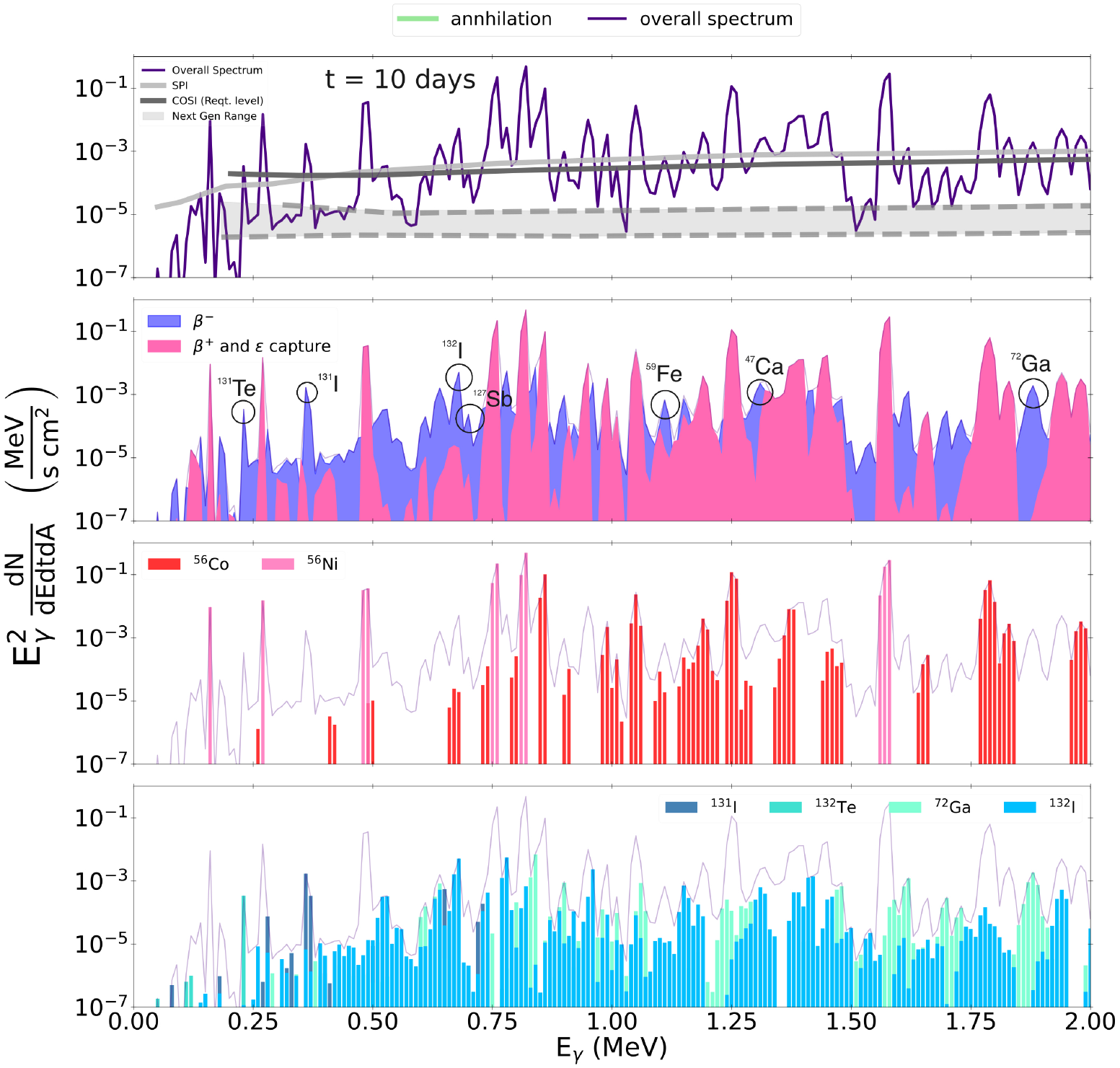}
    \caption{Theoretically calculated gamma-ray spectra of the MR-SN model shown in Fig. \ref{fig:snapshot}, assuming a location of 10 kpc at time $t = 10$ days post explosion. Instrument sensitivities are also scaled to 10 days observing time. The first (top) panel shows the overall spectra and instrument sensitivity limits.  
    The second panel breaks down the total spectrum into the contributions from different nuclear decays: beta minus and the combination of beta plus and $\epsilon$ capture. Although the light curve is dominated by $^{56}$Ni and $^{56}$Co gammas from beta plus decay at this early time, some $r$-process isotopes, such as $^{131}$I, $^{132}$I, $^{132}$Te dominate in a few energy windows. 
    The third and fourth panels indicate prominent lines for each type of decay.}
    \label{fig:10days}
\end{figure}

To represent distinct phases of the early stage emission, we select snapshots at 10 days, 100 days, and 6 years after the explosion, assuming a source distance of 10 kpc. The  spectra are displayed in Figs.~\ref{fig:10days}, \ref{fig:100days} and \ref{fig:6years} using multiple panels. The top panel in each figure presents the total spectrum, overlaid with detector sensitivity curves from selected instruments. Since these spectra are denser than those for the remnants, we use continuum instrument sensitivities as discussed in Sec. \ref{sec:instrument}.
The second panel separates the contributions from beta plus decay isotopes and beta minus decaying isotopes. The subsequent panels isolate prominent gamma lines and their associated isotopes. 
In the remainder of this section we 
discuss key features of these spectra. 
%
%
%

\subsection{10 Days}
Figure~\ref{fig:10days} presents the gamma-ray spectrum 10 days after the explosion, along with the continuum sensitivity limit scaled to 10 days.  On the top panel, we see that the gamma-ray flux is strong at that time, and a large set of features is well above the sensitivity of all instruments.  At this early stage, the ejecta remain optically thick, so the features stem from emission relatively close to the outer edge of the supernova.  From the second panel, we see that the highest peaks are from beta plus decay, and from the third panel we see that these beta plus decays are from $^{56}$Ni and $^{56}$Co, consistent with  Fig.~\ref{fig:lc}. 

A more detailed look at the second panel compared with the first shows that in some energy regions, gammas from beta minus decay are also above telescope sensitivity and not obscured by the more abundant gammas from beta plus decay.  We see, for example, that the second peak $r$-process element $^{132}{\rm I}$, the daughter product of $^{132}$Te with a half-life of 3 days,  has a line at 667 keV that is above the sensitivity limit for all instruments.  The isotope $^{131}{\rm I}$, with a half-life of 8 days, has a similarly prominent line at 364 keV, as does $^{132}$Te with its 228 keV line. Detecting any of these lines would constitute the discovery of an $r$-process that proceeded to the second peak.  We also see a prominent 1861 keV line from near the first $r$-process peak, $^{72}$Ga. This would also constitute an exciting discovery, but it would not necessarily hint at the explosion mechanism because  $^{72}$Ga may also be present in neutrino/advection-driven CCSN if the CCSN produces a weak $r$-process.  Finally, we note that there are also a few additional beta-minus lines visible in the second panel that are above the next generation detector sensitivity including $^{127}{\rm Sb}$ at 685 keV, $^{59}{\rm Fe}$ at 1099 keV and $^{47}{\rm Ca}$ at 1297 keV.
 
The lime green region marks the estimated contribution from e$^{+}$e$^{-}$ annihilation, including both para- and ortho-positronium components, with $^{56}$Co being the primary source of positrons at this time. The emission of 511 keV radiation is above the sensitivity limit for all instruments, and the extended emission tail due to the three-gamma positronium channel is above the sensitivity limit for next-generation detectors.  We assume that spectral features due to nuclear decays that fall below the extended annihilation contribution are effectively obscured and unlikely to be observable.  However, the 228 keV, 364 keV and 667 keV lines that would indicate $r$-process could be identified above this background.

\subsection{100 Days}

%
%
%

\begin{figure}
    \centering
    \includegraphics[width=1\linewidth]{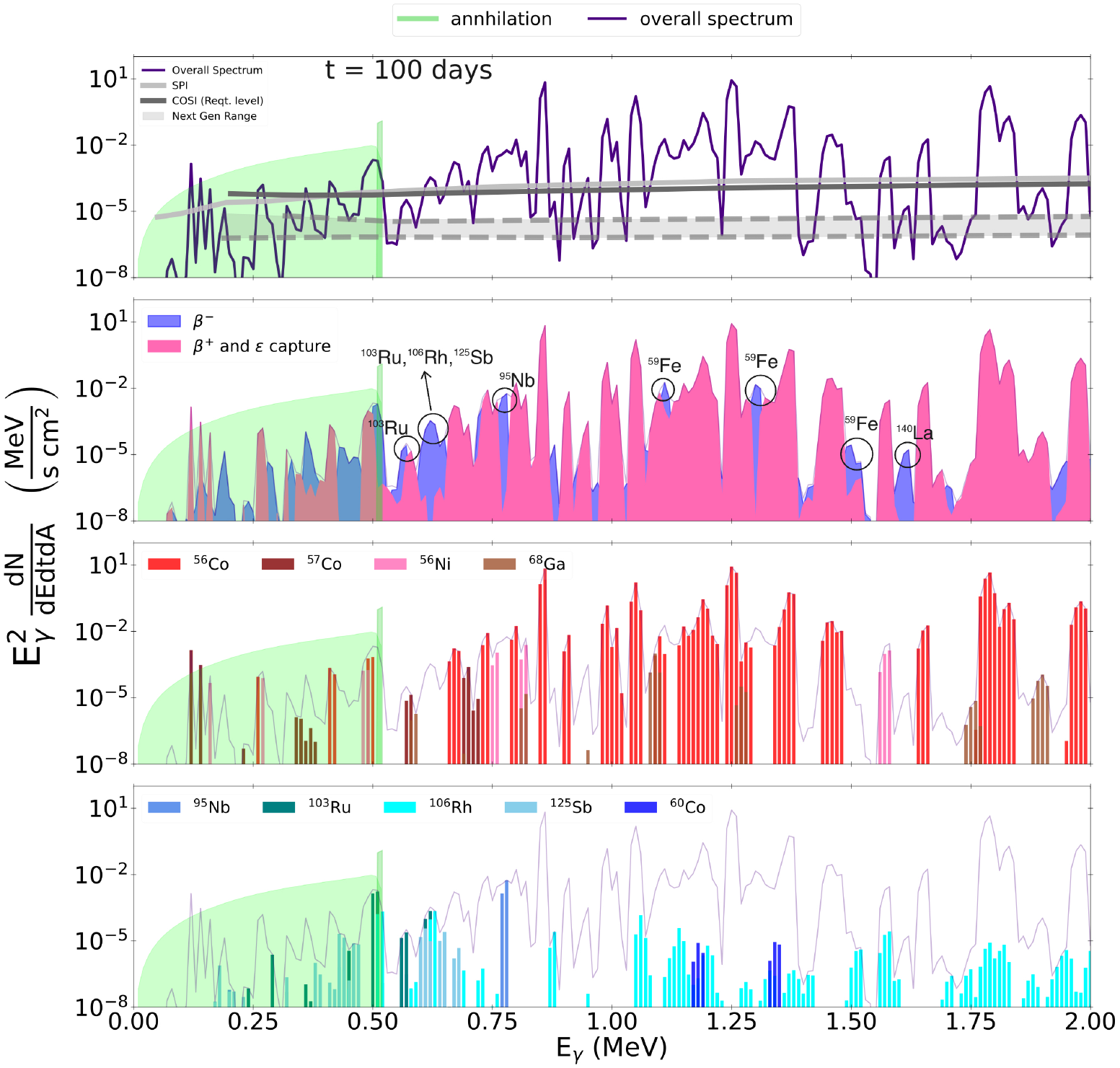}
    \caption{Spectra of the MR-SN at 10 kpc at 100 days post explosion.  At this time, the decays of $^{56}$Ni and $^{56}$Co are still significant, and the annihilation contribution from para-positronium and ortho-positronium is strong. 
    However, between 0.55 MeV and 0.75 MeV, there is a window for weak $r$-process isotopes such as $^{103}$Ru, $^{106}$Rh, $^{125}$Sb, and $^{95}$Nb. The style is the same as Fig. \ref{fig:10days}, although telescope sensitivities are scaled to 100 days of observation time. }
    \label{fig:100days}
\end{figure}

Figure~\ref{fig:100days} presents the gamma-ray spectrum 100 days after the explosion along with the continuum sensitivity limit scaled to 100 days. As illustrated in the left panel of Fig.~\ref{fig:lc}, at this stage the light curve from nuclear gamma emission is dominated by 
$^{56}$Co. In addition, by this time, the cumulative decay of $^{56}$Co generates a pronounced e$^{+}$e$^{-}$ annihilation signal, which likely obscures the majority of gamma-ray lines below 511 keV. Despite this, several heavy isotopes emerge as potential contributors to the spectrum above 511 keV. In particular, a feature around 620 keV is generated by the 621 keV line of $^{106}$Rh, the 600 keV of $^{125}$Sb, and the 610 keV line of $^{103}$Ru. This feature is above the sensitivity limit for next-generation detectors. Importantly, $^{125}$Sb has a relatively long half-life ($t_{1/2} \approx 2.76$ years) and
its lines will remain strong over 100 days, enhancing its detectability.
The isotope $^{95}{\rm Nb}$, with a half-life of $t_{1/2} \approx 35$ days, has a line at 765 keV  (99\% decay intensity) between the decay features from $^{56}$Ni and $^{56}$Co. The isotope $^{103}{\rm Ru}$ has one line at 497 keV that is obscured by the 511 keV radiation, but also a smaller line at 557 keV that is above the next-generation instrument sensitivity.  Finally, we note that there are a few other beta minus decays of interest, including the 1099 keV, 1291 keV, and 1481 keV lines of $^{59}{\rm Fe}$, and the 1596 keV line of $^{140}{\rm La}$.  The isotope $^{140}{\rm La}$ is a short-lived daughter product of $^{140}{\rm Ba}$ that has a half-life of 13 days, and it has become visible now because much of the $^{72}{\rm Ga}$ that obscured it at ten days has decayed.

\subsection{6 Years}
\begin{figure}
    \centering
    \includegraphics[width=1\linewidth]{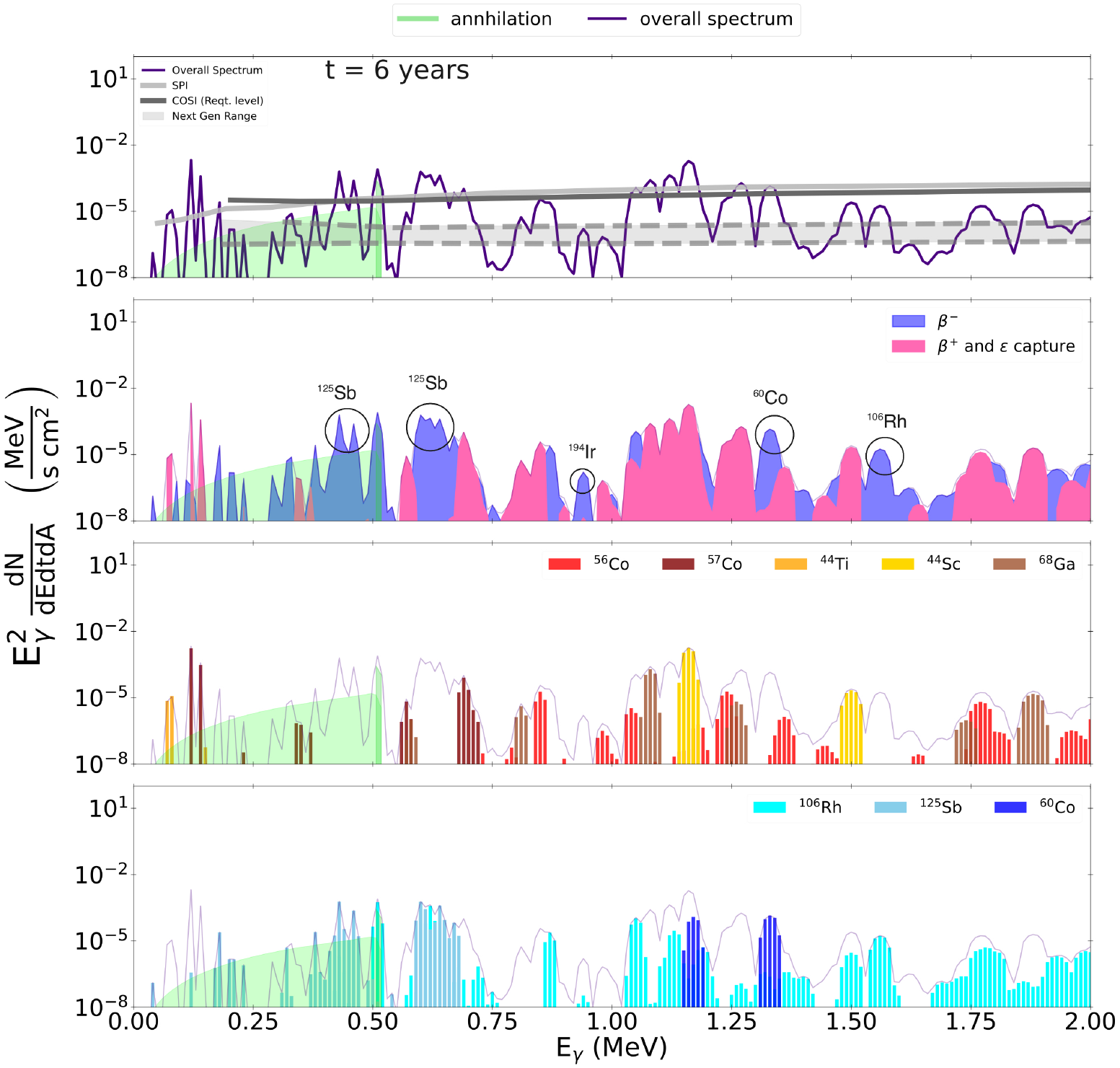}
    \caption{Spectra of the MR-SN at 10 kpc at 6 years after the explosion.  The energy windows for the $r$-process lines are significantly expanded, and the isotopes $^{125}$Sb and $^{106}$Rh dominate the region between 0.55 MeV and 0.75 MeV. The contributions from $^{56}$Ni and $^{56}$Co are significantly reduced at this time due to their relatively short half-lives, and the annihilation contribution primarily stems from positrons produced in the beta plus decay of $^{44}$Sc.  Telescope sensitivities are scaled to 1 year of observation time. }
    \label{fig:6years}
\end{figure}

Figure~\ref{fig:6years} shows the prompt gamma-ray spectrum at 2200 days (approximately 6 years post explosion), with the telescope sensitivity scaled to a one-year observation time. By this stage, the beta plus decay signals from $^{56}{\rm Ni}$ and $^{56}{\rm Co}$ are significantly reduced. From the left panel of Fig. \ref{fig:lc}, we see that the dominant isotope for the MeV light curve is now $^{57}{\rm Co}$ and from Fig.~\ref{fig:6years} we see that it has its two most prominent lines at 122 keV and 136 keV. 

The primary source of positrons that participate in the annihilation radiation at this time originates from the beta plus decay of $^{44}$Sc.  The 511 keV line is above instrument sensitivities, but the weakened annihilation signal compared to 100 days means that the region below 511 keV opens up, and the beta-decaying nuclei are no longer obscured.  For example, in the lower energy range, lines from  $^{44}$Ti (67.8 keV and 78.3 keV) are above this background.  Importantly,  the near-second peak isotope $^{125}{\rm Sb}$, which would be a diagnostic of second peak $r$-process production, dominates the 400 keV region, featuring prominent lines at 427 keV and 463 keV that rise above both the background of annihilation and all the instruments' sensitivities. In our model we find  the \(^{125}\)Sb mass to be \(8.9 \times 10^{-5} M_\odot\) which is lower than the yields from \citet{2017ApJ...836L..21N} but higher than the yields in \cite{2015ApJ...810..109N} and higher than a subset of the models in \cite{2021MNRAS.501.5733R}.  

In the higher energy region, the isotope $^{125}{\rm Sb}$ also contributes several lines in the region from 600 keV to 700 keV which are above the sensitivities of all instruments.  As discussed in Section \ref{sec:remnant}, $^{60}$Co contributes two prominent lines at 1173 keV and 1332 keV, although the 1173 keV line lies close to the 1157 keV line from $^{44}$Sc, complicating its detection. However, the 1332 keV line remains well-separated and above next-generation detector sensitivities.  
We note that there is a line at 938 keV from the decay of $^{194}{\rm Ir}$, which is the short-lived daughter product of the longer-lived (with a half-life of 6 years) $^{194}{\rm Os}$, which might be visible by next-generation detectors.  This is more difficult to detect but is of interest because it is a third-peak $r$-process isotope.  A future detection or upper limit of the mass of $^{194}{\rm Os}$ would be important to understand how far the rapid neutron capture proceeded in the supernova. Finally, the peak in the region of 1562 keV produced by $^{106}{\rm Rh}$ is also well separated and observable by next-generation detectors.

%
%
%

%
%

\section{Conclusion}

Modern simulations of standard, neutrino-driven explosions do not support $r$-process production, although theoretical models indicate that rare types of CCSNe, such as MR-SNe or collapsars, could produce neutron-rich environments conducive to $r$-process nucleosynthesis. Currently, we lack direct observational evidence to confirm or refute the presence of $r$-process elements in supernovae. Gamma-ray observatories offer an opportunity to probe this important question; the observation of gamma-ray radiation from beta decays of newly synthesized unstable $r$-process isotopes would be a direct discovery of $r$-process formation in a supernova.  Alternatively, the absence of such radiation could place a limit on the amount of $r$-process element production.    
 
In this study we consider a specific model, MR-SN - 35OC-Rs from \cite{2021MNRAS.501.5733R}, as an example of $r$-process producing supernovae and use this to evaluate the gamma-ray flux.  Our findings suggest that if an $r$-process producing supernova event were to occur in our Galaxy, multiple gamma-ray lines would be above next-generation gamma-ray telescope sensitivities, such as GRAMS and AMEGO, at distances of $d \sim 10$~ kpc, and some would be above COSI sensitivities. At times less than roughly 1000~days, the gamma-ray signal is dominated by the decays of $^{56}$Ni and $^{56}$Co, as is also expected in a standard CCSN.  However, there are some energy regions where the iron peak isotopes would not conceal signals from $r$-process elements.  For example, at 10 days post explosion, $^{132}{\rm I}$, $^{132}{\rm Te}$ and $^{131}{\rm I}$ all have prominent lines in the 200 keV to 700 keV region which lie both above the annihilation flux and above COSI sensitivity.  Their detection would indicate an $r$-process which proceeded at least to the second peak.  At 100 days post-explosion, the region below 511 keV is mostly obscured by the annihilation flux, but above this energy region there are prominent $r$-process lines from the weak $r$-process elements  $^{106}{\rm Rh}$ and $^{95}{\rm Nb}$ as well as the second peak element $^{125}{\rm Sn}$. A determination of the relative strengths of these lines would address a critical question of the ratio of weak $r$-process to second-peak $r$-process.  

At 6 years post explosion, there are energy regions where the $r$-process isotopes $^{106}{\rm Rh}$, $^{125}$Sb and the neutron-rich $^{60}{\rm Co}$ dominate the nuclear flux and rise above next-generation sensitivity.
The isotope $^{44}{\rm Ti}$ comes primarily from explosive burning and therefore will have similar yields in both CCSNe and MR-SNe, producing lines from the decays of $^{44}{\rm Ti}$ and $^{44}{\rm Sc}$. In addition, the beta plus decay of $^{44}{\rm Ti}$ produces a positron that can annihilate with an electron within the ejecta, creating a peak at 511 keV and a tail at lower energy.   Observations of these would shed new light on the tension between the theoretical predictions of $^{44}{\rm Ti}$ and its observation.

An alternative to observing gamma lines from a prompt explosion is to examine supernova remnants. Roughly eleven supernova remnants are close enough so that gamma radiation from $r$-process radioisotopes would be above instrument sensitivities. Of particular interest are the gamma-ray lines of $^{126}{\rm Sb}$, a beta minus decay daughter of the long-lived isotope $^{126}$Sn.  With a half-life $t_{1/2} \approx 2 \times 10^5$ years, most of the $^{126}$Sn produced in the $r$-process has not substantially decayed over the lifetime of most known CCSN remnants. In the context of the MR-SN model we study, three of the remnants are situated within COSI's sensitivity range, while eight remnants are located within the range of the next-generation telescopes. The most prominent decay lines for $^{126}$Sb are at 666~keV, 695~keV, and 720~keV, and an observation of these lines by COSI or a next-generation gamma-ray telescope would indicate an $r$-process that proceeded to at least the second peak.

In MR-SNe, the most striking features in the supernova remnant spectra come from 
$^{60}$Co, which has prominent gamma-ray lines at 1173~keV and 1332~keV. It derives from the beta minus decay of long-lived $^{60}{\rm Fe}$, which in our model has a yield that is three orders of magnitude higher than the yield of $^{126}{\rm Sb}$.  
While $^{60}{\rm Fe}$ is also present in neutrino/advection driven CCSN ejecta, it is present in much larger quantities if there is a substantial component of neutron-rich outflow, e.g. in the jets of MR-SNe.  Therefore, an observational estimate of the amount of $^{60}{\rm Fe}$ would suggest one scenario or the other.  However, even for the case where the amount of $^{60}{\rm Fe}$ favored MR-SNe, the detection of $^{60}{\rm Co}$  ($^{60}{\rm Fe}$) alone would not guarantee coproduction with $r$-process elements.

In summary, our work explored the potential of detecting the gamma-ray lines from specific isotopes as a method for identifying $r$-process-producing supernovae, such as MR-SNe. We found that multiple lines of $r$-process isotopes are above the detection sensitivities for COSI or next-generation MeV detectors.   A positive detection would represent a paradigm-shifting discovery for CCSN theory. Within the framework of current theoretical models, the detection of gamma-ray lines from second-peak $r$-process isotopes is suggestive of an energy ejection mechanism driven by magnetic fields. However, it would also indicate the need to re-evaluate the role of neutrino-driven CCSNe in $r$-process nucleosynthesis and to further examine collapsars. Such a detection also would recommend a revision of the Galactic chemical evolution models to account for the contributions of rare CCSNe to heavy-element enrichment. In contrast, observational upper limits on such lines would constrain the amount of second-peak and weak $r$-process material that could have been produced in the observed SN.

The feasibility of detecting elements of the $r$-process depends on their production levels, observational sensitivities, and advances in instrumentation. Future improvements in theoretical modeling, including refined magnetohydrodynamic simulations and updated nuclear reaction data, are necessary to predict $r$-process signatures with greater accuracy. COSI and the next generation gamma-ray observatories, such as AMEGO, eAstrogam, MeVGRO, and GRAMS, present the opportunity to further our understanding of Galactic heavy element formation. 

\section*{Acknowledgments}
 LZH would like to thank Dr. Stephen Reynolds for the many hours of excellent and fruitful conversations about supernova remnants. XW would like to thank Dongya Guo and Wenxi Peng for helpful discussions on MeV detector sensitivity. Also, we would like to thank Almudena Arcones, Milena Crnogorcevic, Erika Holmbeck, and Tim Linden for valuable discussions. This work was partially supported by the Office of Defense Nuclear Nonproliferation Research and Development (DNN R\&D), National Nuclear Security Administration, U.S. Department of Energy (GCM, LZH, RS) under contract number LA22-ML-DE-FOA-2440. 
We acknowledge support from the U.S. Department of Energy contract Nos.  
DE-FG0202ER41216 (EG, GCM), DE-FG0295ER40934 (RS), and DE-SC00268442 (ENAF - GCM, EG, RS), as well as from the NSF (N3AS PFC) grant No. PHY-2020275 (GCM, KL, RS), and NSF grant AST~2205847 (IUR). KL, GCM and RS, thank the Institute for Nuclear Theory for its kind hospitality and stimulating research environment, U.S. Department of Energy grant No. DE-FG02- 00ER41132. This work benefited from the travel support provided to GCM, KL, LZH, and RS  from the National Science Foundation Grant No. OISE-1927130 (IReNA). M.R. acknowledges support from the
grants FJC2021-046688-I and PID2021-127495NB-I00, funded by MCIN/AEI/10.13039/501100011033 and by the European Union ``NextGenerationEU” as well as ``ESF Investing in your future”. Additionally, he acknowledges support from the Astrophysics and High Energy Physics program of the Generalitat Valenciana ASFAE/2022/026 funded by MCIN and the European Union NextGenerationEU (PRTR-C17.I1).
The work of XW is supported by the National Key R$\&$D Program of China (2021YFA0718500), the National Natural Science Foundation of China (Grant No. 12494574) and the Chinese Academy of Sciences (Grant No. E329A6M1).

\bibliography{citation,MRbibliography}{}
\bibliographystyle{aasjournal}

\appendix
\section{Positron Annihilation}
\label{app:annihilation}

In this appendix, we perform an order-of-magnitude estimate of the probability of positron energy loss and annihilation. We begin with the electron number density 
\begin{equation}
    n_{e} = \frac{M_{\text{ej}}}{\frac{4}{3}\pi R_{\text{max}}^{3}}N_{A} Y_e,
    \label{eq:electrondensity}
\end{equation}
where $M_{ej}$ is the mass of the ejecta, $R_{\text{max}}$ is the radius of ejecta, $N_{A}$ is Avogadro's constant, and we take the electron fraction to be $Y_e \approx 0.5$, assuming the environment is fully ionized. Equation \ref{eq:electrondensity} yields $n_{e} \approx 10 $ cm$^{-3}$ for Cassiopeia A and $n_{e} \approx 10^{4}$ cm$^{-3}$ for SN1987A.  
For comparison, previous studies of Cassiopeia A yield an $n_{e}$ upper limit of $100 $ cm$^{-3}$ \citep{2009ApJ...693..713S} and a range of $1$ cm$^{-3}$ to $10$ cm$^{-3}$  \citep{2014ApJ...785....7D,2018A&A...612A.110A}. For SN1987A, estimates of $n_{e}$ range from $10^{3}$ cm$^{-3}$ to $10^{6}$ cm$^{-3}$ depending on the ejecta type\citep{2019ApJ...886..147L}.  

We assume that the primary energy loss mechanism for positrons is Coulomb scattering with free electrons at rest. The positron energy loss rate and annihilation probability are calculated using the Bethe-Bloch formula as in\citet{1979tpa..book.....G}:
\begin{equation}
    \left(\frac{dE}{dt}\right)_{\rm COU} = -7.7 \times 10^{-9}\,\frac{n_{e}}{\beta}\left[\ln\left(\frac{E}{n_{e} m_ec^2}\right) + 73.6\right] \quad \text{eV/s}
    \label{equation:energy_lost}
\end{equation}
where $E$ is the energy of the positron in units of eV, $\beta = v/c$, $v$ is the velocity of the positron, $m_e$ is the mass of the electron, and $c$ is the speed of light. In this expression, $n_e$ has units of ${\rm cm}^{-3}$.

Using Equation \ref{equation:energy_lost}, we can get the average distance $\ell$ that a positron with initial energy $E_{i}$ will travel before reaching its final energy $E_{f}$ in a uniform medium with constant $n_e$ from the following:
\begin{equation}
    \ell=\int_{E_i}^{E_f}\frac{d\ell}{dE} dE= \int_{E_i}^{E_f}\frac{d\ell}{dt} \frac{dt}{dE} dE =  3.9 \times 10^{18} \int_{E_i}^{E_f}\frac{\left(1-\frac{m_{e}^2c^4}{E^2}\right)}{n_e\left(\ln\left(\frac{E}{n_em_ec^2}\right)+73.6\right)} dE \quad  \text{cm}.
    \label{equation:l}
\end{equation} 
Positrons emitted by beta plus decay lose much of their initial kinetic energy, typically around 0.6 MeV for decays of \({}^{56}\)Co and \({}^{44}\)Sc \footnote{Calculated from ENSDF database available at http://www.nndc.bnl.gov/ensarchivals/}, through ionization and scattering before annihilating in flight with free electrons (producing two 511 keV photons) or forming positronium. So here we set $E_{i}\simeq (511 + 600)\, {\rm keV}$ for the positrons emitted by the beta plus decays $^{56}$Ni, $^{56}$Co, and $^{44}$Sc. We take the final energy to be the rest energy of the positron, $E_{f} = 511\,{\rm keV}$, to calculate the maximum distance that an emitted positron can travel with a given $n_e$, which is defined as $\ell (n_e)$.

\begin{figure}
    \centering    \includegraphics[width=0.3\linewidth]{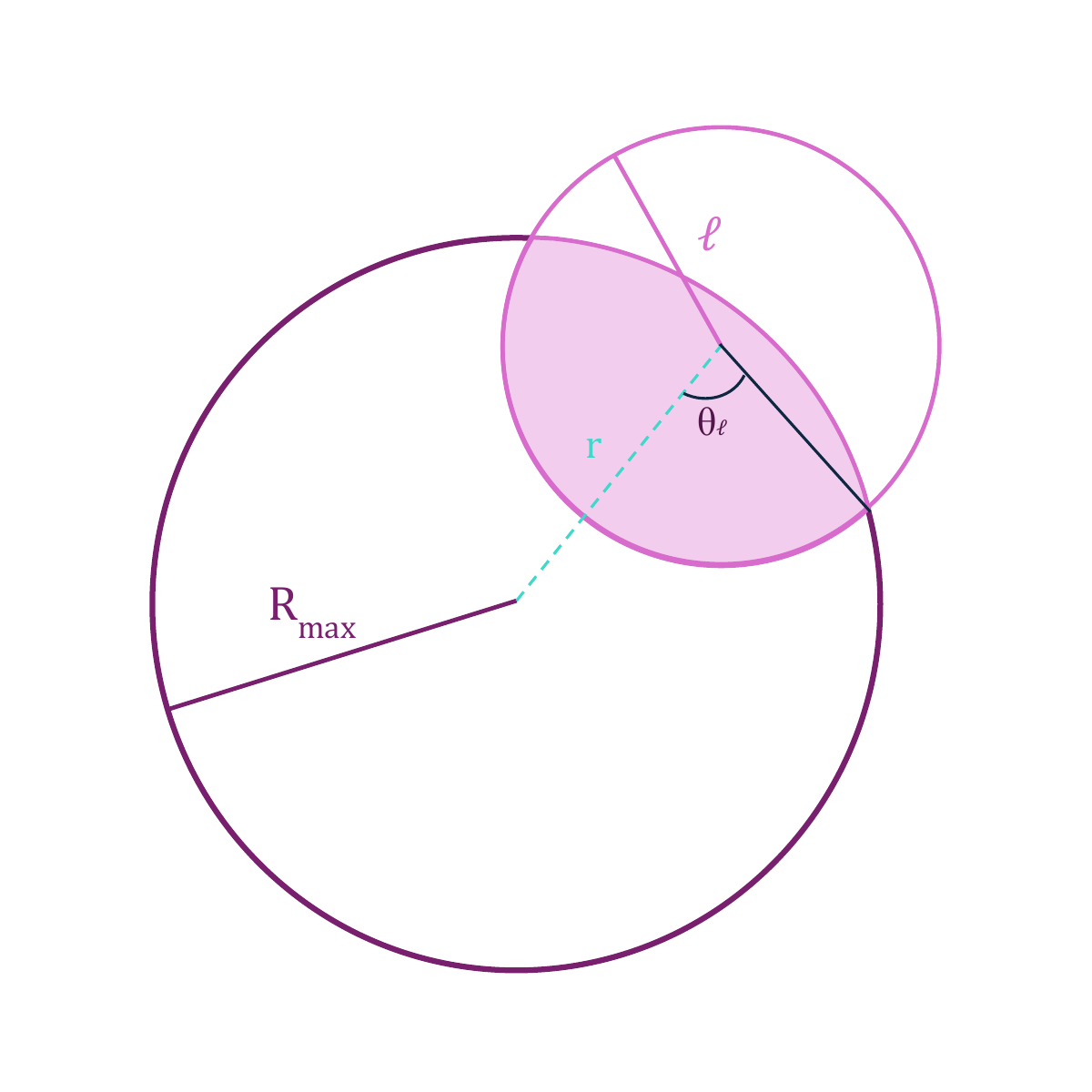}
    \caption{Sketch of the geometric setup used to estimate the positron annihilation signal. The ejecta is represented as a sphere of radius \( R_{\text{max}} \), shown as the dark purple circle. At a radial coordinate \( r \), we define an auxiliary \(\ell\)-sphere with radius \(\ell\) and elevation angle \(\theta\), shown as the magenta circle. The two spheres intersect at the elevation angle \(\theta_\ell\), beyond which positrons would escape the ejecta. The solid angle subtended by the \(\ell\)-sphere for \(\theta < \theta_\ell\) corresponds to the trajectories of positrons that come to rest and annihilate within the ejecta. This region is illustrated with medium purple shading. 
}
    \label{fig:anni_Drawn}
\end{figure} 

  We calculate $\ell$ in Equation \ref{equation:l} assuming that the positron stays within the ejecta for its entire trajectory.  However, in reality, this will not be the case for some trajectories.  To find the fraction of positrons that reach $E_f$ before escaping, we calculate the overlap of the surface of the sphere with radius $\ell$ centered at the emission point $r$, with the sphere of expanding ejecta with radius $R_{\text{max}}$. See Fig. \ref{fig:anni_Drawn} for a sketch of $\ell$ and $R_{\text{max}}$. The $\ell$-sphere surface is completely, partially, or not at all contained within the ejecta sphere, depending on the values of $r$ and $\ell$.
For the portion of the $\ell$-sphere contained within the ejecta sphere, we assume that the positron will annihilate inside the ejecta sphere with a probability of $100\%$.
For the remainder of the $\ell$-sphere exterior to the ejecta sphere, we assume that the positron will escape.  

 If $\ell<2R_{\text{max}}$, then we take into account all possible emission radii by integrating over the entire ejecta sphere ($0 < r < R_{\text{max}}$).  For each radius, we integrate over the included solid angle of the $\ell$-sphere to calculate the total annihilation probability $P_{\rm ann}(\ell)$
\begin{subequations}
\begin{align}
    P_{\rm ann}(\ell) &= \frac{1}{4 \pi V} \left( \int_{0}^{R_{\text{max}}-\ell}4\pi r^{2}\,dr\int_{-1}^{1}2\pi \,d\cos\theta + \int_{R_{\text{max}}-\ell}^{R_{\text{max}}}4\pi r^{2}\,dr\int_{\cos\theta_{\ell}}^{1}2\pi \,d\cos\theta \right) & (\ell < R_{\text{max}});
    \label{equation: Pannl<R}\\
    P_{\rm ann}(\ell) &= \frac{1}{4 \pi V} \left(\int_{\ell-R_{\text{max}}}^{R_{\text{max}}}4\pi r^{2}\,dr\int_{\cos \theta_{\ell}}^{1}2\pi \,d\cos\theta \right) & (R_{\text{max}}<\ell<2R_{\text{max}})
    \label{equation: Pannl>R}
\end{align}
\end{subequations}
Since we are describing a probability, we need to normalize the number annihilating by the number produced, so we divide by the total solid angle $4 \pi$ and the ejecta volume $V$.
The angle $\theta_l$ describes where the $\ell$-sphere and ejecta sphere intersect:
\begin{equation}
    \cos \theta_{\ell} = \frac{r^2+\ell^2-R_{\text{max}}^2}{2r\ell}
\end{equation}
For the spherical model we use, the integrals in Eqs.\ \eqref{equation: Pannl<R} and \eqref{equation: Pannl>R} can be reduced to an analytical expression:
\begin{align}
  P_{\rm ann}(\ell) = 1 - \frac{\ell}{16R_{\text{max}}}\left[12-\left(\frac{\ell}{R_{\text{max}}}\right)^2\right]  \qquad 
  (0<\ell<2R_\text{max}).
  \label{eq:annprob1}
\end{align}
Once $\ell$ becomes larger than $2R_{\text{max}}$ (about 38 years after explosion in our model), we assume that all positrons escape from the ejecta and $P_{ann} \approx 0$. In principle, there will also be in-flight annihilation that will increase the probability beyond what is given in Eq. \ref{eq:annprob1}. Using expressions in  \citet{Prantzos2010} for the in-flight annihilation probability for positrons moving through a medium of electrons, we found that when integrated over the positron trajectories, the net probability of in-flight annihilation is small ($<$ 0.01). 
Therefore, we neglect this effect in our estimates. 

\section{Radiative transfer calculations}
\label{app:radiationtransport}

For radiative transfer calculations, we consider two variants of spherical homologous expansion to model the expanding ejecta.  In the first case, we assume the complete mixing of all ejecta components with a uniform density function in the following form
\begin{equation}
\rho = \frac{3 M_{\text{ej}}}{4 \pi R_{\text{max}}^3}, 
\label{eq:density}
\end{equation}
where $M_{\text{ej}}$ is the total mass ejected and the radius of the sphere at time $t$ is  $R_{\text{max}} =R_{\text{max},i}+v_{\text{ej}}t$. Guided by Fig. \ref{fig:snapshot}, we take the initial radius of the sphere to be  $R_{\text{max},i} = 2 \times 10^9$ cm.  In the main text, we refer to this model as the \lq uniform mix\rq. 

In second case, we consider all explosive burning material to be uniformly mixed and interior to $R_{\text{max},i} = R_{\text{explosive},i} =2 \times 10^9$~cm with the same functional form as Eq. \ref{eq:density} with $M_{\text{ej}}$  replaced by the total mass that undergoes explosive burning $M_{\text{explosive}}$. In this case, the uniformly mixed jet material constitutes an outer layer between $R_{i, \text{explosive}}$ and $R_{i, \text{jet}} = 2.2 \times 10^9$ cm.  Starting at $r = R_{i, \text{explosive}}$, the uniform density of this outer layer is given by: 
\begin{equation}
\rho_{\text{layer}} = \frac{3 M_{\text{jet}}}{4 \pi (R_{\text{jet}}^3-R_{\text{explosive}}^3)}, 
\label{eq:density-layer}
\end{equation}
where $M_{\text{jet}}$ is the total mass of the jet material, and ${R_{\text{max}}= R_{\text{jet}} =R_{i,\text{jet}}+v_{\rm ej}t}$, ${R_{\text{explosive}}=R_{i,\text{explosive}}+v(R_{\text{explosive}})t}$.  
In both cases, the expansion speed is taken at all times to be $v(r) = v_{\text{ej}} r/R_{\text{max}}$ with the outmost layer speed being $v_{\text{ej}} = 0.015c$ \citep{2011piim.book.....D,2017hsn..book.1981R}. 

At late times in both cases, typically after $\sim$3 to 4 years, the ejecta becomes optically thin, allowing gamma-ray photons to escape with minimal interaction. In this regime, the spectrum is effectively adjusted only by Doppler broadening, reflecting the velocity distribution of the radioactive material.
The Doppler broadening can be approximated using the relativistic Doppler formula for the small velocity $\vec{v}$ limit:
\begin{equation}
    E =  E_{\text{rest}} (1 + \frac{\vec{v} \cdot \ \hat{r}_{ls}}{c})
    \label{eq:doppler}
\end{equation}
where $\hat{r}_{ls}$ is the unit vector along the line of sight.
The minimum possible energy of the broadened gammas is $E_{\text{shifted min}} = E_{\text{rest min}} (1 - v_{\text{ej}}/c)$ and the maximum is $E_{\text{shifted max}} = E_{\text{rest max}} (1 + v_{\text{ej}}/c)$.  
The Doppler-shifted spectrum at each location is found by appropriately applying Eq. \ref{eq:doppler} to each part of the outflow, $j$, and redistributing the energy spectrum of the prompt emission 
\begin{equation}
\frac{dN^{\text{prompt}}_j(\vec{r},t)}{dE_{rest}dt} (E_{rest}) \rightarrow \frac{dN_j( \vec{r},t)}{dEdt} (E).
\end{equation} In the optically thin limit, the total contribution for each energy can be computed by integrating over each emitting volume element, taking into account the appropriate direction of the outflow  at each time  
and obtaining the flux by dividing by $ 4 \pi d^2$ where $d$ is the distance from the Earth to the supernova.

In the optically thick epoch, 
we also need to take into account the reduction of gamma-ray photons along the line of sight due to scattering and absorption.   We use the semi-analytical 
method described in \cite{2019MNRAS.486.2910W} and \cite{2020ApJ...903L...3W} to solve the radiative transfer equation $dI_{E}/dl = -\alpha(E)I_{E}+J_{E}$ to get $I_{E}(l)$, by tracking the emission and removal (through absorption or scattering) of gamma-ray photons as they travel through the ejecta. The scattered photons are ignored here. In this equation, $I_{E}$ denotes the energy-specific line intensity which is integrated over the path $l$, in units of photon counts per time per area per energy per solid angle.  The path is taken to be along the line of sight that connects the ejecta center and the observer. The absorption coefficient is taken as $\alpha(E) = \rho \kappa(E)$, where $\kappa(E)$ is the opacity experienced by the photons as they travel through the ejecta. Opacity values are adopted from XCOM\footnote{https://www.nist.gov/pml/xcom-photon-cross-sections-database} for Compton scattering, Rayleigh scattering, photoelectric absorption, and pair production.  We calculate the total opacities of the 
ejecta based on its composition with a combination of the opacity values of the
 characteristic isotopes (Fe, Xe, Eu, Pt, and U) as discussed in \cite{2020ApJ...903L...3W}.
The source term \( J_E \) represents  gamma-ray emissivity with dimensions of photon counts per energy per volume per solid angle. For the "uniform mixing" case, this is:
\begin{equation}
    J_E = \frac{1}{4\pi V_{\text{tot}}} \sum_j \frac{dN_j (E)}{dE\,dt}
\end{equation}
and total volume is given by \( V_{\text{tot}} = 4\pi R_{\text{max}}^3/{3} \).  For the second more structured case, the source term is decomposed into two distinct regions corresponding to explosive burning and jet material:
\begin{equation}
J_E(r) =
\begin{cases}
\displaystyle \frac{1}{4\pi V_{\text{explosive}}} \sum_{j_{\text{explosive}}} \frac{dN_{j_{\text{explosive}}}(E)}{dE\,dt}, &  r < R_{\text{explosive}} \\
\\[-1em]
\displaystyle \frac{1}{4\pi V_{\text{jet}}} \sum_{j_{\text{jet}}} \frac{dN_{j_{\text{jet}}}(E)}{dE\,dt}, &  R_{\text{explosive}} < r < R_{\text{jet}}
\end{cases}
\end{equation}
In this expression, the indices \( j_{\text{explosive}} \) and \( j_{\text{jet}} \) refer to trajectories associated with explosive burning material and jet outflow, respectively. The corresponding volumes are defined as \( V_{\text{explosive}} = 4\pi R_{\text{explosive}}^3/3\) and \( V_{\text{jet}} = 4\pi\left( R_{\text{jet}}^3 - R_{\text{explosive}}^3 \right)/3 \).

Finally, to obtain the total flux in units of photon counts per energy per area per time from the ejecta, the intensity $I_{E}(l)$ is integrated across the entire solid angle subtended by the spherical ejecta as viewed from the detector:
\begin{equation}
    F_{E} = \int I_{E} (\theta)\cos\theta d\Omega
\label{equation:final_flux}
\end{equation}
where $\theta$ denotes the angle between the line of sight and the line between the ejecta center and the observer. Here, the energy $E$  is the observed photon energy after Doppler shift.

\section{Detector Sensitivities}
\label{app:sens}
In this appendix, we compile the $3 \sigma$ significance continuum sensitivity limits from various MeV instruments for gamma-ray energies between 0.2 MeV and 2 MeV and scale to one year of observing time. We also list the line sensitivity limits with $3 \sigma$ significance as published for the corresponding MeV mission.

\begin{table}[htbp]
\centering
\label{tab:contsens}
\footnotesize
\begin{tabular}{|c|c|c|c|c|c|}
\hline
\shortstack{Energy\\(MeV)} &
\shortstack{INTEGRAL/SPI\\sensitivity\\(MeV/cm$^2$/s)} &
\shortstack{COSI\\sensitivity\\(MeV/cm$^2$/s)} &
\shortstack{e-ASTROGAM\\sensitivity\\(MeV/cm$^2$/s)} &
\shortstack{GRAMS\\sensitivity\\(MeV/cm$^2$/s)} &
\shortstack{AMEGO\\sensitivity\\(MeV/cm$^2$/s)} \\
\hline
0.20 & $1.3 \times 10^{-5}$ &  &  & $3.2 \times 10^{-7}$ & $3.0 \times 10^{-6}$ \\
0.40 & $2.8 \times 10^{-5}$ & $2.9 \times 10^{-5}$ & $1.0 \times 10^{-6}$ & $3.6 \times 10^{-7}$ & $3.0 \times 10^{-6}$ \\
0.60 & $5.1 \times 10^{-5}$ & $3.3 \times 10^{-5}$ & $1.0 \times 10^{-6}$ & $3.6 \times 10^{-7}$ & $2.0 \times 10^{-6}$ \\
0.80 & $7.3 \times 10^{-5}$ & $4.1 \times 10^{-5}$ & $2.0 \times 10^{-6}$ & $3.5 \times 10^{-7}$ & $2.0 \times 10^{-6}$ \\
1.00 & $9.5 \times 10^{-5}$ & $4.9 \times 10^{-5}$ & $2.0 \times 10^{-6}$ & $3.5 \times 10^{-7}$ & $2.0 \times 10^{-6}$ \\
1.20 & $1.2 \times 10^{-4}$ & $5.8 \times 10^{-5}$ & $2.0 \times 10^{-6}$ & $3.7 \times 10^{-7}$ & $2.0 \times 10^{-6}$ \\
1.40 & $1.3 \times 10^{-4}$ & $6.6 \times 10^{-5}$ & $2.0 \times 10^{-6}$ & $3.8 \times 10^{-7}$ & $3.0 \times 10^{-6}$ \\
1.50 & $1.4 \times 10^{-4}$ & $7.1 \times 10^{-5}$ & $2.0 \times 10^{-6}$ & $3.9 \times 10^{-7}$ & $3.0 \times 10^{-6}$ \\
2.00 & $1.7 \times 10^{-4}$ & $9.2 \times 10^{-5}$ & $3.0 \times 10^{-6}$ & $4.4 \times 10^{-7}$ & $4.0 \times 10^{-6}$ \\
\hline
\end{tabular}
\caption{Interpolated Continuum sensitivities for multiple telescopes between 0.2 and 2 MeV. Each detector's sensitivity is scaled to 1-year observation time. SPI and e-ASTROGAM sensitivities are adopted from Fig. 1 of \cite{2018JHEAp..19....1D}. GRAMS sensitivity is adopted from \cite{2020APh...114..107A}. COSI sensitivity is adopted from Fig. 2 of \cite{Tomsick:2023aue}. AMEGO sensitivity is adopted from Fig. 5 of \cite{McEnery2019All}.}

\end{table}

\begin{table}[htbp]
\centering
\label{tab:linesens}

{\footnotesize
\begin{tabular}{|c|c|c|c|c|c|}
\hline
\shortstack{Energy\\(MeV)} &
\shortstack{INTEGRAL/SPI\\sensitivity\\(ph/cm$^2$/s)} &
\shortstack{COSI$^{a}$\\sensitivity\\(ph/cm$^2$/s)} &
\shortstack{e-ASTROGAM\\sensitivity\\(ph/cm$^2$/s)} &
\shortstack{GRAMS\\sensitivity\\(ph/cm$^2$/s)} &
\shortstack{AMEGO$^{b}$\\sensitivity\\(ph/cm$^2$/s)} \\
\hline
0.511 (e$^{+}$) & $5.2 \times 10^{-5}$ & $1.2 \times 10^{-5}$ 
& $4.1 \times 10^{-6}$ 
& $6.3 \times 10^{-7}$ &  \\
0.666/0.695 ($^{126}$Sn) & & &  
& $4.2 \times 10^{-7}$ &  \\
0.847 ($^{56}$Co) & $2.3 \times 10^{-4}$ & 
& $3.5 \times 10^{-6}$ 
& $2.7 \times 10^{-7}$ &  \\
1.157 ($^{44}$Ti) & $9.6 \times 10^{-5}$ & $3.0 \times 10^{-6}$ 
& $3.6 \times 10^{-6}$ 
& $1.9 \times 10^{-7}$ &  \\
1.173 ($^{60}$Fe) & & $3.0 \times 10^{-6}$ 
&   
& $1.9 \times 10^{-7}$ &  \\
1.275 ($^{22}$Na) & $1.1 \times 10^{-4}$ & 
& $3.8 \times 10^{-6}$ 
&  &  \\
1.333 ($^{60}$Fe) &  & $3.0 \times 10^{-6}$  
&  
& $1.7 \times 10^{-7}$ &  \\
1.809 ($^{26}$Al) & $2.5 \times 10^{-5}$  & $3.0 \times 10^{-6}$ 
&  
& $1.3 \times 10^{-7}$ & $1 \times 10^{-6}$ \\
2.223 ($^{2}$H) & $1.1 \times 10^{-4}$ & 
& $2.1 \times 10^{-6}$ 
& $1.1 \times 10^{-7}$ &  \\
4.438 ($^{12}$C$^{*}$) & $1.1 \times 10^{-4}$ & 
& $1.7 \times 10^{-6}$ 
& $7.3 \times 10^{-7}$ &  \\
\hline
\end{tabular}
}
\caption{Published $3\sigma$ narrow line sensitivities for example MeV instruments adopted in this paper. INTEGRAL/SPI line sensitivity is adopted from \cite{2003A&A...411L..91R} and \cite{2018JHEAp..19....1D}, e-ASTROGAM line sensitivity is adopted from Table 1.3.3 of \cite{2018JHEAp..19....1D}.
 The COSI narrow line sensitivity is the requirement level adopted from Table 1 of \cite{Tomsick:2023aue}, the AMEGO narrow line sensitivity is adopted from Table 1 of \cite{McEnery2019All}, and the GRAMS narrow line sensitivity is adopted from Table 1 of \cite{2020APh...114..107A}. Notes: a. COSI line sensitivity has 2 years of observation time. b. AMEGO line sensitivity is for a 5 year mission. All other sensitivities have an observation time of $10^{6}$ seconds.}
\end{table}

\newpage

\section{$^{60}{\rm Fe}$ Yields}
\label{sec:60Feyields}
In this appendix, we provide a table of $^{60}{\rm Fe}$ yields from a variety of models published in previous works.
%
%
\begin{table}[h!]
\centering
\begin{tabular}{llll}
\hline
$^{60}{\rm Fe}$ yields ($10^{-3}$ M$_{\odot}$) & Progenitor Type & Model description &Reference \\ \hline
$2.3 \times 10^{-7}$  & CCSN  & [Fe/H] = -3, zero rotation, 25M$_{\odot}$ progenitor &\cite{2018Limongi} \\
$1.2 \times 10^{-5}$  & CCSN  & [Fe/H] = -2 zero rotation, 25M$_{\odot}$ progenitor &\cite{2018Limongi} \\
$2.5 \times 10^{-3}$  & CCSN  & [Fe/H] = -1 zero rotation, 25M$_{\odot}$ progenitor &\cite{2018Limongi} \\
$3.6 \times 10^{-2}$  & CCSN  & W18 engine 25.2M$_{\odot}$ progenitor &\cite{2016Sukhbold} \\
$3.7 \times 10^{-2}$  & CCSN  & [Fe/H] = 0 zero rotation, 25M$_{\odot}$ progenitor &\cite{2018Limongi} \\
$4.0 \times 10^{-1}$  & CCSN  & upper limit of S16 model &\cite{2024Burrows} \\
$4.1 \times 10^{-1}$  & CCSN  & 25M$_{\odot}$ Max &\cite{2020andrews} \\
$1.7 \times 10^{-4}$  & MR-SN & model 35OC-RRw &\cite{2021MNRAS.501.5733R} \\
$1.2 \times 10^{-2}$  & MR-SN & $L_{\nu}$= 0.1 &\cite{2017ApJ...836L..21N} \\
$5.0 \times 10^{-2}$  & MR-SN & $L_{\nu}$= 0.2 &\cite{2017ApJ...836L..21N} \\
$1.6 \times 10^{-1}$  & MR-SN & model 35OC-Rw &\cite{2021MNRAS.501.5733R} \\
$2.8 \times 10^{-1}$  & MR-SN & $L_{\nu}$= 0.4 &\cite{2017ApJ...836L..21N} \\
$4.9 \times 10^{-1}$  & MR-SN & model 35OC-RO &\cite{2021MNRAS.501.5733R} \\
$6.8 \times 10^{-1}$  & MR-SN & B11tw025 &\cite{2015ApJ...810..109N} \\
$7.5 \times 10^{-1}$  & MR-SN & B11tw100 &\cite{2015ApJ...810..109N} \\
$9.9 \times 10^{-1}$  & MR-SN & $L_{\nu}$= 0.3 &\cite{2017ApJ...836L..21N} \\
$2.0 $   & MR-SN & $L_{\nu}$= 0.5 &\cite{2017ApJ...836L..21N} \\
$2.2 $   & MR-SN & B12tw025 &\cite{2015ApJ...810..109N} \\
$2.3 $   & MR-SN & B12tw100 &\cite{2015ApJ...810..109N} \\
$2.7 $   & MR-SN & $L_{\nu}$= 0.6 &\cite{2017ApJ...836L..21N} \\
$4.3 $   & MR-SN & B12tw400 &\cite{2015ApJ...810..109N} \\
$4.4-8.3 \footnote{$4 \times10^{-3}$ M$_{\odot}$ is given in \cite{2021MNRAS.501.5733R} by considering only jet material ejected by the end of the simulation time. $8.3 \times10^{-3}$ M$_{\odot}$ is estimated in this paper after the inclusion of the 0.3 M$_{\odot}$ of the infalling material, which is anticipated to be ejected in the jet, as discussed in Sec \ref{sec:gammas}. } $   & MR-SN & model 35OC-Rs &\cite{2021MNRAS.501.5733R} \\
$5.0 $   & MR-SN & $L_{\nu}$= 0.75 &\cite{2017ApJ...836L..21N} \\
$5.3 $   & MR-SN & $L_{\nu}$= 1.0 &\cite{2017ApJ...836L..21N} \\
$5.7 $   & MR-SN & $L_{\nu}$= 1.25 &\cite{2017ApJ...836L..21N} \\
\hline
\end{tabular}
\caption{Theoretical $^{60}$Fe yield predicted by different MR-SNe and CCSNe studies. The values listed for \cite{2015ApJ...810..109N,2017ApJ...836L..21N} are estimated from the final yields of $^{60}{\rm Ni}$ given in the corresponding references.} 
\label{tab:60Fe_yields}
\end{table}

\end{document}